\documentclass[fleqn,usenatbib]{mnras}

\usepackage{newtxtext,newtxmath}

\usepackage[T1]{fontenc}
\usepackage{pdflscape}
\usepackage{siunitx} 

\DeclareRobustCommand{\VAN}[3]{#2}
\let\VANthebibliography\thebibliography
\def\thebibliography{\DeclareRobustCommand{\VAN}[3]{##3}\VANthebibliography}

\usepackage{graphicx}	
\usepackage{amsmath}	
\usepackage[skip=0pt, indent = 9pt]{parskip}
\usepackage{tablefootnote}

\title[The invisible black widow PSR J1720-0534]{The invisible black widow PSR J1720-0534: implications for the electron density towards the North Polar Spur}

\author[K. I. I. Koljonen et al.]{
K. I. I. Koljonen,$^{1}$\thanks{E-mail: karri.koljonen@ntnu.no}
S. S. Lindseth,$^{1}$
M. Linares$^{1,2}$ 
A. K. Harding$^{3}$
and M. Turchetta$^{1}$
\\
$^{1}$Department of Physics, Norwegian University of Science and Technology, NO-7491 Trondheim, Norway\\
$^2$Departament de F{\'i}sica, EEBE, Universitat Polit{\`e}cnica de Catalunya, Av. Eduard Maristany 16, E-08019 Barcelona, Spain.\\
$^{3}$Theoretical Division, Los Alamos National Laboratory, Los Alamos, NM 87545 }

\date{Accepted XXX. Received YYY; in original form ZZZ}

\pubyear{2023}

\begin{document}
\label{firstpage}
\pagerange{\pageref{firstpage}--\pageref{lastpage}}
\maketitle

\begin{abstract}
Radio emission from pulsars can be used to map out their distances through dispersion measure (DM), which quantifies the amount of radio pulse dispersion. However, this method relies on accurately modelling the free electron density in the line of sight. Here, we present a detailed study of the multiwavelength emission from PSR J1720$-$0534, a black widow compact binary millisecond pulsar discovered in 2021, which the latest electron density model of the Galaxy \citep{yao17} places at only 191 pc. We obtained and analysed deep multiwavelength observations in the $\gamma$-ray (Fermi-Large Area Telescope, 2008–2022), optical (Las Cumbres Observatory, 2.7 h), near-infrared (Nordic Optical Telescope, 3.5 h), and X-ray (Swift-X-Ray Telescope, 10 ks) bands. We found no significant detection of $\gamma$-ray, optical, near-infrared, or X-ray counterparts around the radio-timing position of PSR J1720$-$0534, which we thus nickname `the invisible black widow'. Employing the most constraining near-infrared limit ($J>23.4$ mag), we established a lower limit on the source distance, $d>1.1$ kpc, assuming conservative properties for the black widow companion star. This distance lower limit differs drastically (by a factor of more than 5) from the Yao et al. DM distance estimate. We attribute this difference to the inclusion in the Yao et al. model of a large and dense component towards the North Polar Spur. Considering our results and recent parallax distances to other pulsars in this direction, we argue that such a local and large component in the electron density model of the Galaxy is unnecessary.

\end{abstract}

\begin{keywords}
 Galaxy: local interstellar matter -- pulsars: general -- pulsars: individual: PSR J1720-0534 -- stars: distances -- stars: neutron
\end{keywords}



\section{Introduction}\label{sec:intro}

The radio sky outside the Galactic plane exhibits several prominent emission features, with large radio loops dominating the high-latitude sky \citep[e.g.,][]{berkhuijsen71,haslam71}. The North Polar Spur (NPS) stands out as the most significant among these features \citep{hanburybrown60}. This structure has been recognized as part of the Loop I configuration known since the 1950s \citep{baldwin55}. The NPS is a particularly bright streak of radio emission perpendicular to the Galactic plane, spanning from $l=30^{\circ}$, $b=0^{\circ}$ up to Galactic latitudes of $b=40^{\circ}$. The prevailing explanation for the existence of these large loops involves one or more supernova remnants that originated in close proximity to the Solar system, within a few tens of parsecs, which occurred relatively recently, within the past $10^5-10^6$ years \citep[e.g.,][]{spoelstra73,salter83}.

Several radio filaments exist within Loop I towards the Galactic Centre, and not all of them are thought to be associated with nearby structures. A notable example is the Fermi bubbles, which form a double-lobed feature centered on the Galactic Centre, detected at gamma-ray energies \citep{su10} and microwave wavelengths \citep{dobler10}. The northern bubble extends to a latitude of $b=55^{\circ}$ and has a width of $17^{\circ}$ in longitude. The likely origin of the Fermi bubbles is attributed to either relatively recent active galactic nucleus type activity at the Galactic Centre involving a remnant of a jet ejection event or a bipolar Galactic wind from starburst activity \citep{su10}. In addition, recent studies have questioned the local origin of NPS/Loop I \citep{predehl20,iwashita23}, with some earlier works also suggesting a distant origin \citep[e.g.,][]{sofue00}, pointing towards much larger, Galactic-scale structures. From our viewpoint inside the Galaxy, a mixture of nearby and Galactocentric radio structures can likely explain the produced radio loops and filaments \citep{lallement23}.   

PSR J1720$-$0534 is an eclipsing compact binary millisecond pulsar (CBMP) located at the Galactic coordinates $l\sim17^{\circ}$ and $b\sim17^{\circ}$ in the direction of NPS/Loop I/Fermi Bubbles, and was discovered in 2021 using the Chinese Five-hundred-meter Aperture Spherical Telescope (FAST). \citet{wang21} and \citet{miao23} reported a rotational period of $P=3.26$\,ms for this pulsar orbiting an $M_c \gtrsim 0.034$\,M$_\odot$ brown dwarf companion in a 3.16-h orbit. The light mass of the companion places the source in the CBMP subcategory of black widows \citep[BWs;][]{fruchter88}, which typically have semidegenerate brown dwarf companions with masses of $M_c \gtrsim 0.01$\,M$_\odot$. Using the most recent electron density model of \citet[][hereafter YMW16]{yao17}, PSR J1720$-$0534 was estimated to be at a distance of $d=191$\,pc, making it the nearest BW known so far. However, using the older electron distribution model from \citet[][hereafter NE2001]{ne2001}, the source distance is estimated to be much farther at 1.3 kpc.

In this paper, we present a detailed $\gamma$-ray, near-infrared, optical, and X-ray analysis of the region around PSR J1720$-$0534 (Section 2). We do not find a significant detection of the counterpart in any of the studied wavelengths (Section 3). Our deep limiting near-infrared magnitude can place constraints on the source distance, assuming that the companion star is not atypically small or cold (Section 3.2). We discuss the implication for the large and dense component representing the NPS/Loop I in YMW16 (Section 4.1). We argue that the dispersion measure (DM)  of PSR J1720$-$0534 is mostly due to the Galactic thick disc component, which results in a likely distance of 3.1 kpc (Section 4.2). We conclude and summarize in Section 5.

\section{Observations and Data Analysis}\label{sec:obsanddata}

\subsection{Optical}

We observed PSR J1720$-$0534 using the Multicolor Simultaneous Camera for studying Atmospheres of Transiting exoplanets 3 (MuSCAT3) multichannel optical imager mounted on the 2-m Las Cumbres Observatory (LCO) telescope on 2022 May 28. The exposures, lasting 10 min each, were simultaneously taken using the \textit{g'}, \textit{r'}, \textit{i'} and \textit{z}$_{s}$ camera channels in a $9 \times9$ arcmin$^{2}$ field of view around our target. We gathered a total of $2.7$ h of observations, with a full width at half maximum (FWHM) ranging from 0.9 to 1.6 arcsec across the images.

We processed the data using the \textsc{banzai} data-processing pipeline.\footnote{https://lco.global/documentation/data/BANZAIpipeline/}, which included tasks such as bad-pixel masking, bias subtraction and flat-field correction. Additionally, we combined the images into a single deep image for each of the four optical bands, \textit{g'}, \textit{r'}, \textit{i'} and \textit{z}$_{s}$, to enhance source detection sensitivity.

\subsection{Near-infrared}

We observed PSR J1720$-$0534 with the Nordic Optical Telescope near-infrared Camera and spectrograph (NOTCam) near-infrared instrument at the Nordic Optical Telescope (NOT) on 2023 June 13. We utilized the imaging mode and the $J$-band filter. The observations comprised 24 sequences of nine-point dither images, shifted in a 3$\times$3 grid pattern around the source with a step size of 10 arcsec. Each dither image was taken in ramp-sampling mode, with 10 readouts every 6 s during a 60-s integration time. Thus, the total integration time per single dither sequence was 540 s, resulting in a total observation time of 3.6 h. The FWHM measured in the $J$-band images ranged from 0.54 to 0.8 arcsec over the observation. 

We processed the data using the NOTcam \textsc{quicklook v.2.6} reduction package.\footnote{
http://www.not.iac.es/instruments/notcam/guide/observe.html} The reduction process included creating a differential master flat, implementing linearity corrections and bad pixel masking, performing sky subtraction, and stacking the dithered images. Additionally, we stacked the combined dither images to create a single deep image to search for the infrared counterpart of PSR J1720$-$0534. 

\subsection{X-rays}

We reduced and analysed the X-ray Telescope (XRT) data of PSR J1720$-$0534 from the \textit{Neil Gehrels Swift Observatory} \citep[\textit{Swift}/XRT,][]{burrows05}. We obtained a target-of-opportunity observation on J2023 January 24, with a net exposure of approximately 10 ks. Using \textsc{xselect v2.5} in \textsc{HEASoft 6.31.1}, we extracted an image in the 0.3--10 keV band derived from the cleaned photon-counting mode event data (obtained by running \textsc{xrtpipeline v0.13.6}). Source detection in the generated image was performed using the \textsc{detect} algorithm with no significantly detected sources in the image. The upper limit estimation was carried out using the \textsc{sosta} algorithm in the  \textsc{HEASoft}'s X-ray image analysis package \textsc{Ximage v4.5.1} at the location of PSR J1720$-$0534 using source half-box size of 5 pixels and rectangular background annulus with the inner half-box size of 20 pixels and outer half-box size 38 pixels. The resulting 3-$\sigma$ upper limit is 0.0015 cts s$^{-1}$.

\subsection{$\gamma$-rays}

We conducted a search for continuous $\gamma$-ray emission from PSR J1720$-$0534 using \textit{Fermi}/Large Area Telescope (LAT) data covering the time range from 2008 August 4 to 2022 September 26. To achieve this, we utilized the Pass 8 SOURCE class of \textit{Fermi}/LAT events, focusing on the energy range of $1-1000$\,GeV, with a maximum zenith angle of $z=100^\circ$. The data were centred on the position of the extended \textit{Fermi}/LAT source FHES J1723.5$-$0501 (4FGL J1723.5$-$0501e; \citealt{ackermann18}), associated with a Type 1a supernova remnant (SNR; G17.8+16.7; \citealt{araya22}). The central coordinates of this extended source are only $0.85^\circ$ away from the radio location of PSR J1720$-$0534 (see Appendix \ref{sec:extended source}). We included data within $8^\circ$ of the central position of FHES J1723.5-0501 for our analysis. 

A joint likelihood analysis of the data was conducted using \texttt{evtype=32} and \texttt{evtype=28}, where the former represents the set of events in the best quartile of the point spread function (PSF) partition, and the latter is the joint set of the three worst quartiles. The data were binned with an angular pixelation of $0.025^\circ$ and divided into eight energy bins per decade. The configuration of the data selection and analysis is summarized in Table \ref{tab:dataselec}.

To optimize the nearby extended source FHES J1723.5$-$0501, we employed an optimization algorithm based on the work of \citet{ackermann18}. Here, we utilized the Fermi Point Source Catalog-Data Release 3 (4FGL-DR3)\footnote{https://fermi.gsfc.nasa.gov/ssc/data/access/lat/12yr\_catalog} sources \citep{abdollahi22} along with the \texttt{gll\_iem\_v07} and \texttt{iso\_P8R3\_SOURCE\_V3\_v1} background models for the Galactic and isotropic background radiation, respectively. These models are provided by the \textit{Fermi}-LAT Collaboration\footnote{See https://fermi.gsfc.nasa.gov/ssc/data/access/lat/BackgroundModels.html}. 

The spectral parameters of the model sources are fitted using the \textsc{newminuit} $\chi^2$ minimization algorithm \citep{James:1994vla}, which is implemented within the \textsc{Fermi Science Tools} version 2.2.0. To apply the tools to the data, we utilized the \textsc{fermipy python} package version 1.2. Similar to \citet{ackermann18}, we constrained the region of interest (ROI) to a $6^\circ\times6^\circ$ box, while including 4FGL catalogue sources up to a $10^\circ\times10^\circ$ region. However, in contrast to their analysis, we kept the spectral parameters of the sources outside the ROI fixed to their catalogoe values throughout the entire optimization algorithm. Further details of the ROI optimization algorithm can be found in Appendix \ref{sec:extended source}.

\section{Results}

\subsection{Optical}

We searched for the optical counterpart of PSR J1720$-$0534 at its radio location \citep[RA: 17:20:54.506, Dec.: $-$05:34:23.822;][]{miao23} in our combined optical images but did not find any nearby sources (within 7 arcsec). The weakest sources found in the combined optical images have magnitudes $g'=23.9$ mag, $r'=23.3$ mag, $i'=22.7$ mag, and $z_{s}=22.1$ mag. 

Given that the companion stars in BW systems are cool, with non-irradiated nightside effective temperatures ranging from 1000 K to 3000 K \citep{draghis19,matasanchez23,turchetta23}, this positions the peak emission wavelength of the companion star at approximately 1--1.5 $\mu$m ($J$-band). Consequently, we expect the BW companions to be brighter at near-infrared wavelenghts. 

\subsection{Near-infrared}\label{sec:nir}

\begin{figure}
    \centering
    \includegraphics[width = 1.0\columnwidth]{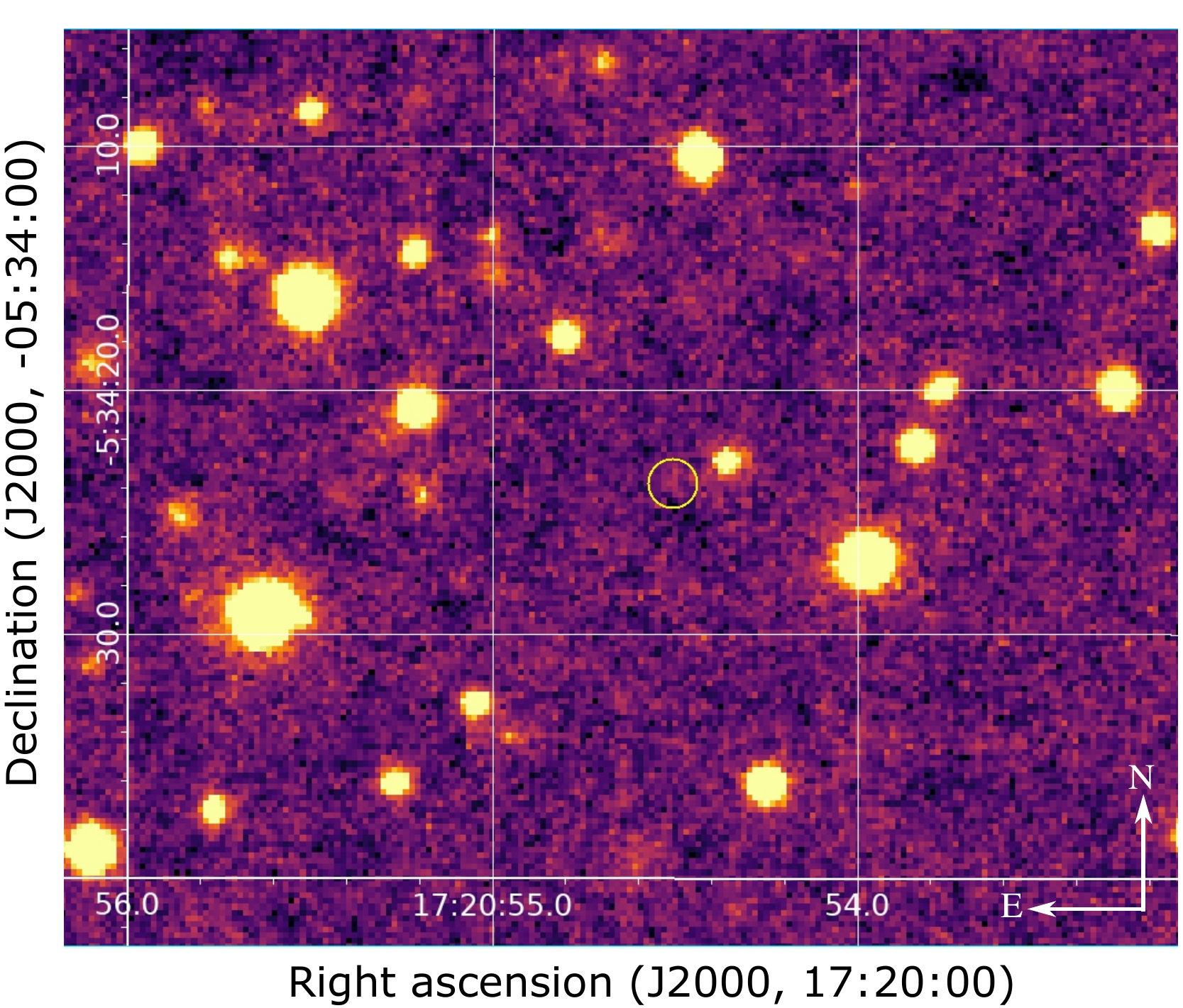}
    \caption{NOTCam $J$-band image with a 3.5-h exposure centred on the radio location of PSR J1720$-$0534 (yellow 1 arcsec circle). We did not detect any source in this region. The angular distance to the nearest source towards the northwest is 2.5 arcsec. However, we do not consider this to be the infrared counterpart of PSR J1720$-$0534 (see text for more details).}
    \label{fig:IR}
\end{figure}

\begin{figure*}
    \centering
    \includegraphics[width = 1.0\columnwidth, angle=270]{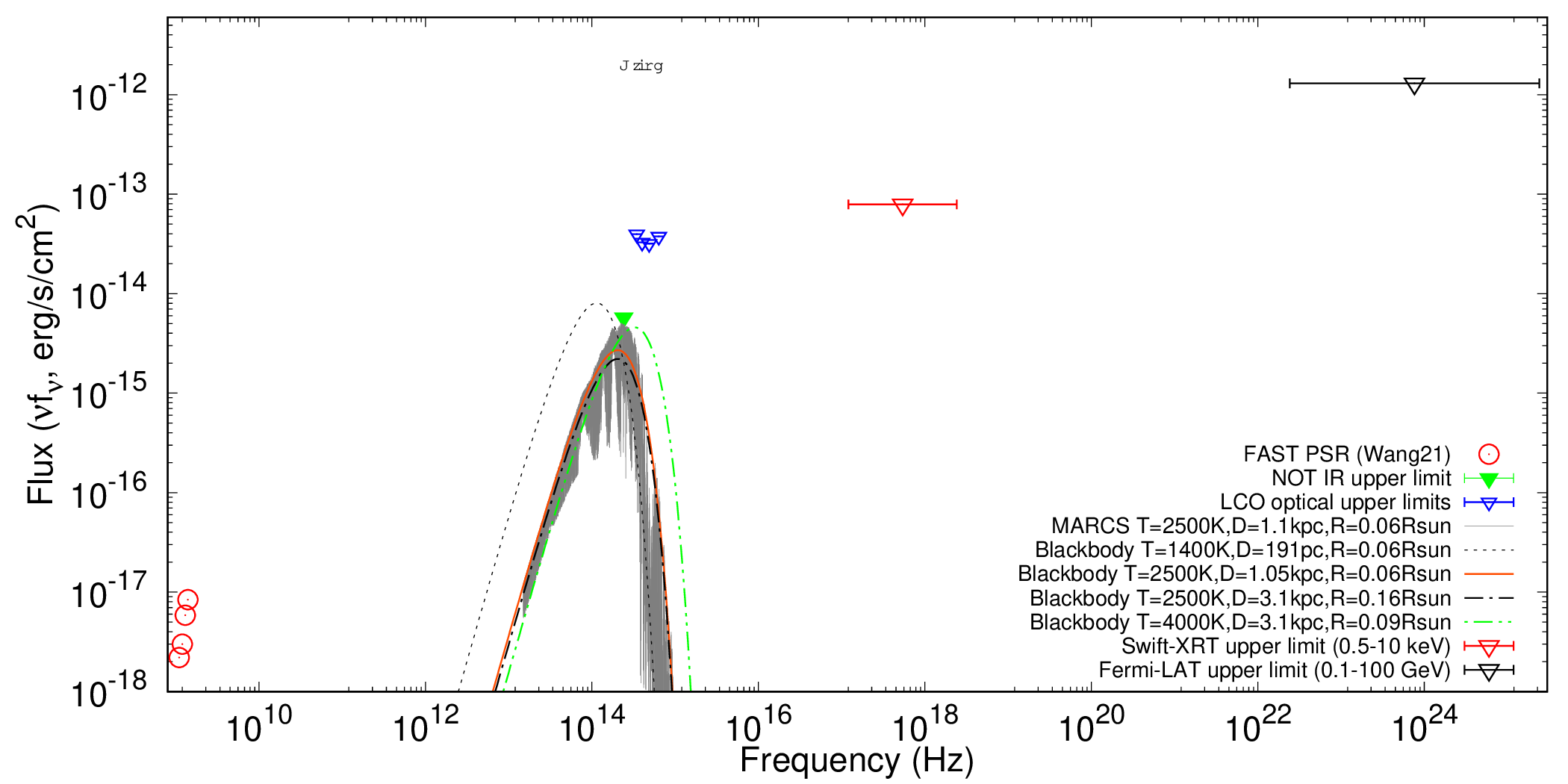}
    \caption{Our upper limits on the infrared ($J$-band; green, filled triangle), optical (z$_{s}$, i'-, r'-, and g'-bands; blue open triangles), X-ray (red, open triangle), and $\gamma$-ray (black, open triangle) fluxes of PSR J1720$-$0534 along with the observed radio fluxes obtained with FAST \citep[red, open circles,][]{wang21}. The grey lines depict the stellar atmosphere model spectrum of a dwarf star with a temperature of 2500 K from MARCS, scaled to the distance of 1.05 kpc and a stellar radius of 0.06 $R_{\odot}$. Additionally, we include several blackbody spectra with varying temperatures, distances, and radii as discussed in the text. Note that the stellar atmosphere model significantly differs from a pure blackbody, especially at the peak.} 
    \label{fig:sed}
\end{figure*}

Surprisingly, no infrared source was detected at the radio location of PSR J1720-0534 (Fig. \ref{fig:IR}) either in the individual $J$-band images or in the combined image. The nearest source to PSR J1720$-$0534 is located 2.5 arcsec north-west, with an apparent $J$-band magnitude of approximately 20.2. However, it does not exhibit significant orbital variability (Appendix \ref{sec:varIR}). 

The time difference between the radio reference epoch and our NOTCam observations is 3.06 yr. If the nearest source had moved from the radio location to the observed one, it would imply a proper motion exceeding 800 mas yr$^{-1}$. While the radio timing solution of \citet{miao23} does not include an estimate for the proper motion, the highest measured proper motion for a pulsar is 375 mas yr$^{-1}$ \citep[PSR B1133+16, ATNF catalog, v. 1.70;][]{brisken02,deller19}. Therefore, given the large angular distance and the absence of orbital variability, we conclude that the nearest source is not the infrared counterpart of PSR J1720$-$0534.

Since the dimmest sources found in the combined $J$-band image have magnitudes up to 23.4, this provides a conservative lower limit for the magnitude of the near-infrared counterpart of PSR J1720$-$0534. To place a lower limit on the distance to the system, we employ the most conservative properties for the companion star from the currently known population: the smallest known radius and the lowest known average temperature of a BW companion; $R \approx 0.06$ $R_{\odot}$ and $T_{\rm eff}\approx2500$ K, respectively \citep{matasanchez23}. Since our observations cover a full orbit and all known BW companions are irradiated (see Section \ref{sec:dist_to_J1720}), we use the average between dayside and nightside temperatures.

We estimated the $J$-band extinction at the location of PSR J1720$-$0534 as $A_{\rm J}$=0.44 utilizing the 3D dust map \textsc{bayestar19} \citep{green19}. We can now place a lower limit on the distance to PSR J1720$-$0534 using a stellar spectrum for the above temperature from the Model Atmospheres with a Radiative and Convective Scheme (MARCS) synthetic stellar spectral library \citep{gustafsson08}\footnote{Specifically, we use one with $T = 2500$ K, $\rm{log}(g)=4.5$, and solar abundances. The lowest available temperature in the library is $T = 2500$ K. While the abundances of the BW companions are relatively little known, first studies show a departure from solar values \citep{shahbaz22}. However, using higher abundances or surface gravities has very little effect on the flux at less than 10 per cent level, and would increase it at near-infrared wavelengths.}, and scaling it according to the minimum radius and a given distance (Fig. \ref{fig:sed}, grey lines). We find that the distance corresponding to our $J$-band limiting magnitude is 1.05 kpc, and thus we can place a conservative distance lower limit to the source as $d_{J} > 1050$ pc.\footnote{In a similar fashion, distance lower limits using the optical limiting magnitudes and appropriate extinction values ($A_{z_s} = 0.85$, $A_{i'} = 1.08$, $A_{r'} = 1.44$, $A_{g'} = 1.93$) can be derived, although they are much less constraining: $d_{g'} > 43$ pc, $d_{r'} > 50$ pc, $d_{i'} > 186$ pc, $d_{z_s} > 274$ pc.} If the companion star fills or is close to filling its Roche Lobe, which is the case in many BW systems \citep[$\sim$60\%;][]{matasanchez23}, with a volume-equivalent radius of $R = 0.16$ $R_{\odot}$ \citep{eggleton} calculated using the orbital parameters from \citet{miao23}, it would correspond to a distance lower limit of $d_{J} > 3.1$ kpc (Fig. \ref{fig:sed}, black dot-dashed line). On the other hand, if the close distance estimate of YMW16 ($d=191$ pc) is the assumed distance, it would require a very low average temperature of $T<1400$ K (Fig. \ref{fig:sed}, black dotted line).

\subsection{X-rays}

We did not detect any X-ray source at the radio location of PSR J1720$-$0534 in the 10 ks \textit{Swift}/XRT image. We can place a conservative 3$\sigma$ upper limit on the source flux in the 0.5--10 keV band using an X-ray power-law photon index of $\Gamma=1.5$ \citep[lowest value among BWs;][]{swihart22} and a line-of-sight Galactic hydrogen column density of $N_{\mathrm{H}} = 1.3\times10^{21}$ atoms/cm$^2$ \citep{HI4PI} resulting in $F_{\mathrm{X}}<8\times10^{-14}$ erg/s/cm$^2$. 

On the other hand, the lowest X-ray efficiency (the ratio of the X-ray luminosity to spin-down power; $\eta \equiv L_{\mathrm{X}}/\dot{E}$) measured for a BW is for PSR J0636$+$5129: $\eta=1.4\times10^{-5}$ \citep[][although we note that this depends on the DM distance that has a large uncertainty]{spiewak16}. Similar minimum efficiencies are measured for other pulsars with thermal X-ray spectra \citep[e.g.,][]{kargaltsev12,posselt12,vahdat22}. Since the spin-down power of PSR J1720$-$0534 is known from radio timing observations \citep[$\dot{E}=9.2\times10^{33}$ erg/s;][]{miao23}, this would imply a minimum X-ray luminosity of $L_{\mathrm{X}}=1.3\times10^{29}$ erg/s. Convolving this with the above flux upper limit places a lower limit for the distance of PSR J1720$-$0534: $d>165$ pc. However, this limit does not include the correction from the Shklovskii effect \citep{shklovskii70} since the proper motion of the source is not known. It also assumes isotropic emission. In any case, because some BWs have very low X-ray luminosity \citep[$L_{\mathrm{X}} \lesssim 10^{30}$ erg/s;][]{swihart22}, this non-detection of PSR J1720$-$0534 is not surprising. 

\subsection{$\gamma$-rays}

\begin{figure}
    \centering
    \includegraphics[width = 1\columnwidth]{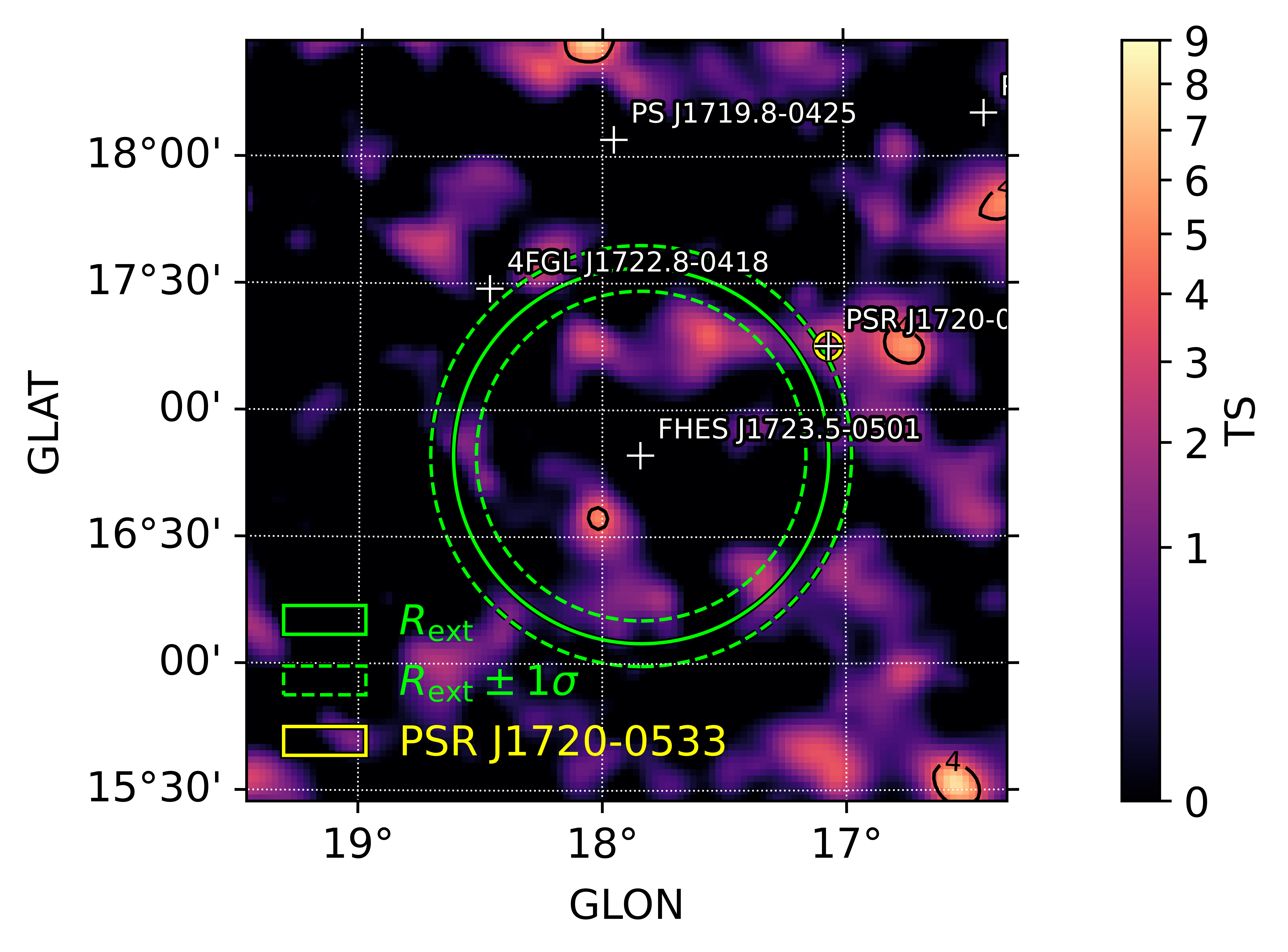}
    \caption{TS map of the region around PSR J1720$-$0534 using a test source with a typical spectral shape of a pulsar. When producing the map, PSR J1720$-$0534 is excluded from the model to show any potential emission around its position. The yellow circle indicates the radio position of PSR J1720$-$0534. The green circles indicate $R_\mathrm{ext}$ of FHES J1723.5$-$0501 with $\pm 1\sigma$ uncertainties (see Appendix \ref{sec:extended source}). The white crosses on the map indicate the positions of the model sources. No significant source is detected at the position of the pulsar.}
    \label{fig:TSpsr}
\end{figure}

After optimizing the ROI with the extended source, we added a point source at the position of PSR J1720$-$0534. We modeled the pulsar's spectral shape with a power law with a super-exponential cut-off (PLSEC) given by

\begin{equation}
    \frac{dN}{dE}=N_0\left(\frac{E}{E_0}\right)^\Gamma \text{exp}\left(-\frac{E}{E_c}\right)^b.
\end{equation}

\noindent This is a common way to model the spectral shapes of millisecond pulsars \citep{abdo13}. For PSR J1720$-$0534, we used the cut-off energy $E_c = 3.7$\,GeV and spectral index $\Gamma=1.54$, which are the best-fitting values reported by \citet{xing16} for a sample of 39 millisecond pulsars from the Second \textit{Fermi}/LAT Catalog for $\gamma$-ray pulsars \citep{abdo13}. We set $b=1$ as there is no evidence to suggest a subexponential cut-off for PSR J1720$-$0534, and we also set the scale factor $E_0$ to $1$\,GeV.

After introducing the point source to the model, we fitted its normalization together with all spectral parameters of all other model sources inside the ROI. From this final fit, we found test statistic (TS) $=2.8$ for PSR J1720$-$0534, resulting in a detection significance of $1.7\sigma$ and thus a non-detection of $\gamma$-rays from the pulsar. Figure \ref{fig:TSpsr} presents the local TS map around the pulsar using a test source with a PLSEC spectral model with $\Gamma=1.54$ and $E_c = 3.7$\,GeV, and the map shows no significant peaks in the vicinity of the pulsar's position. For the TS peak around $0.3^\circ$ away from the pulsar position, we found a maximum TS value of $\mathrm{TS}=5.6$ ($\sigma\approx2.4$), which is not a significant detection either.

For the spectral properties assumed above, we found a 95 per cent upper limit on the $\gamma$-ray energy flux in the $0.1-100$\,GeV band of $G_\gamma < 1.3\times10^{-12}$\,erg\,cm$^{-2}$\,s$^{-1}$. Compared to X-rays, the minimum $\gamma$-ray efficiencies for pulsars are higher, at around 1 per cent level \citep[e.g.,][]{kargaltsev12,smith23}, which for the spin-down power for PSR J1720$-$0534 would mean a minimum $\gamma$-ray luminosity of $L_{\gamma}=9.2\times10^{30}$ erg/s. Convolving this with the above flux upper limit places a lower limit for the distance of PSR J1720$-$0534: $d>240$ pc. Similar to X-rays, this limit does not include the correction from the Shklovskii effect and assumes isotropic emission.    

\section{Discussion}\label{sec:discussion}

The quality of an electron density model is determined by comparing the resulting DM distances to independently measured pulsar distances, typically obtained through parallax measurements. YMW16 represents an enhancement over the NE2001 model by incorporating more recent data and updating information on known systematic effects \citep[e.g.,][]{lorimer06,chatterjee09}. YMW16 also removes small-scale voids and clumps from the electron density model to prevent issues of overfitting. However, it retains certain local structures, such as the Gum Nebula, the Local Bubble, and the NPS/Loop I.

Further comparisons of the electron density models have been conducted by \citet{deller19} and \citet{price21}. \citet{deller19} compared a sample of pulsar distances using radio-timing parallax from the PSR$\pi$ survey, concluding that while YMW16 provide more accurate distances for high-latitude pulsars in the sample, both models do not agree with parallax distances for a few sources, and overall the pulsar distances are underestimated. A similar underestimation was observed when using \textit{Gaia} parallax distances for a sample of CBMPs \citep{koljonen23}. 

The largest differences between the DMs of YMW16 and NE2001 are found at the location of small-scale clumps in NE2001 and at larger features at low Galactic latitudes, such as the Gum Nebula, Local Bubble, or NPS/Loop I, and NE2001’s low-density region \citep{price21}. Notably, two pulsars, PSR J1735$-$0724 and PSR J1741$-$0840 (located close to PSR J1720$-$0534), have particularly poorly estimated DM distances in YMW16. The discrepancy appears to be due to excess electron density within 200 pc, attributed to the contribution of the NPS/Loop I component in the model. This suggests that either the electron density of NPS/Loop I is overestimated and/or that the actual location is different from the modelled one.

In addition, the lower limits on the distance of PSR J1720$-$0534, particularly the one derived from our near-infrared observations, contradict the distance estimate based on the DM using the electron density model of YMW16 (191 pc) at least by a factor of 5. Instead, they align more closely with the distance estimate from NE2001 (1.3 kpc). In the following, we discuss the implications of our distance limits for PSR J1720$-$0534 and several other pulsars with parallax-based distances in the same direction for the electron density models.

\subsection{Is the NPS/Loop I component needed in YMW16?}

\begin{figure*}
    \centering
    \includegraphics[width = 1.0\textwidth]{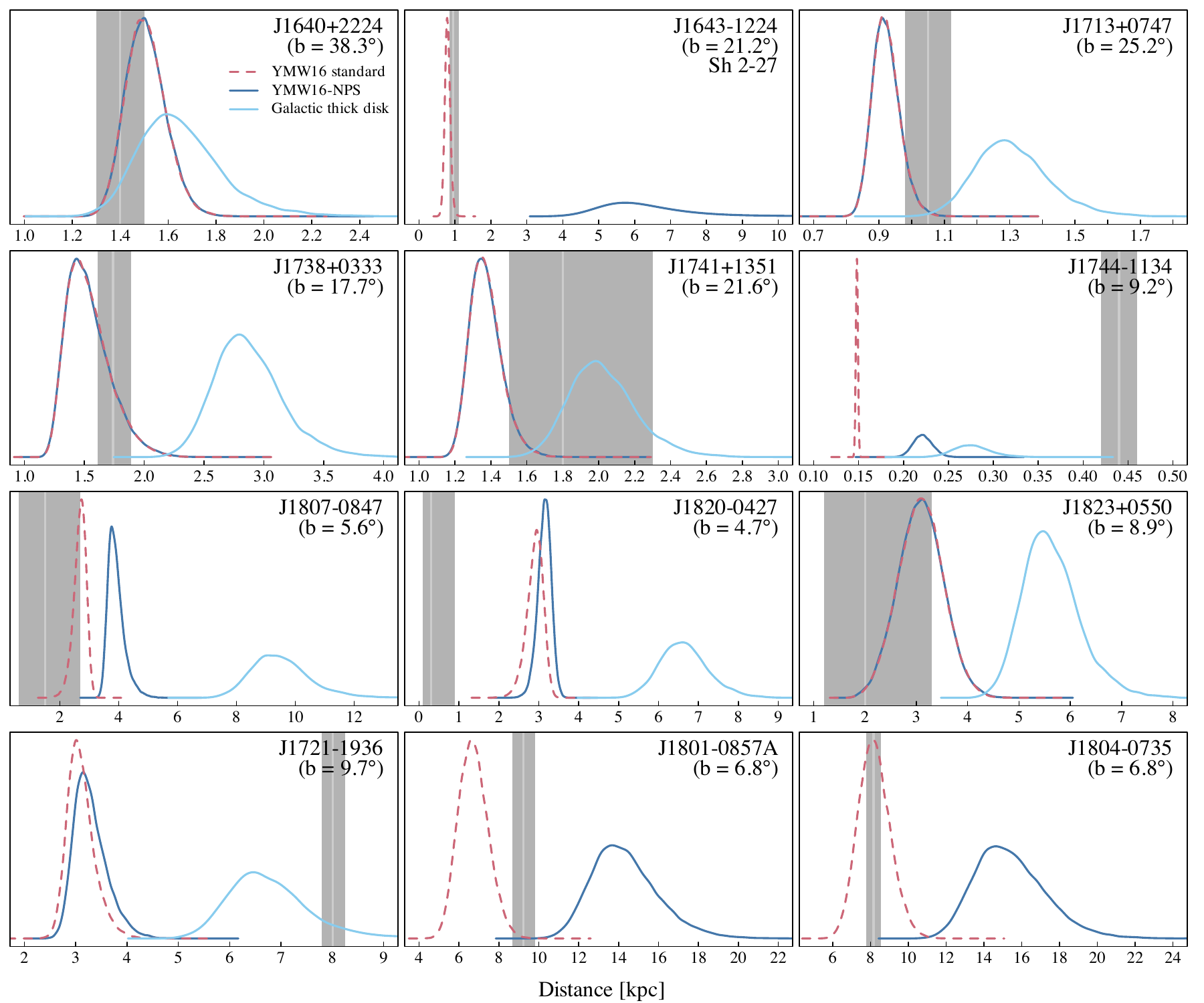}
    \caption{DM distance posteriors of pulsars towards the NPS/Loop I (each panel indicates the pulsar name, its Galactic latitude, and possible association to H \textsc{II} region). Three distance posteriors are shown: YMW16 electron density distribution model with standard parameters (red, dashed lines), excluding the NPS/Loop I component (dark blue, solid lines), and using only the Galactic thick disc component (light blue, solid lines). Note that in cases where the pulsar location does not coincide with the NPS/Loop I component in YMW16, the first two posteriors are identical. For pulsars PSR J1643$-$1224, PSR J1801$-$0857A, and PSR J1804$-$0735, the DM distance using only the Galactic thick disc component is not constrained. The light gray vertical line and dark gray bands around it show the distance measurement and its 1$\sigma$ confidence interval from independent methods (see Table \ref{tab:pulsar_table}).}
    \label{fig:pulsars}
\end{figure*}

\begin{figure}
    \centering
    \includegraphics[width = 1.0\columnwidth]{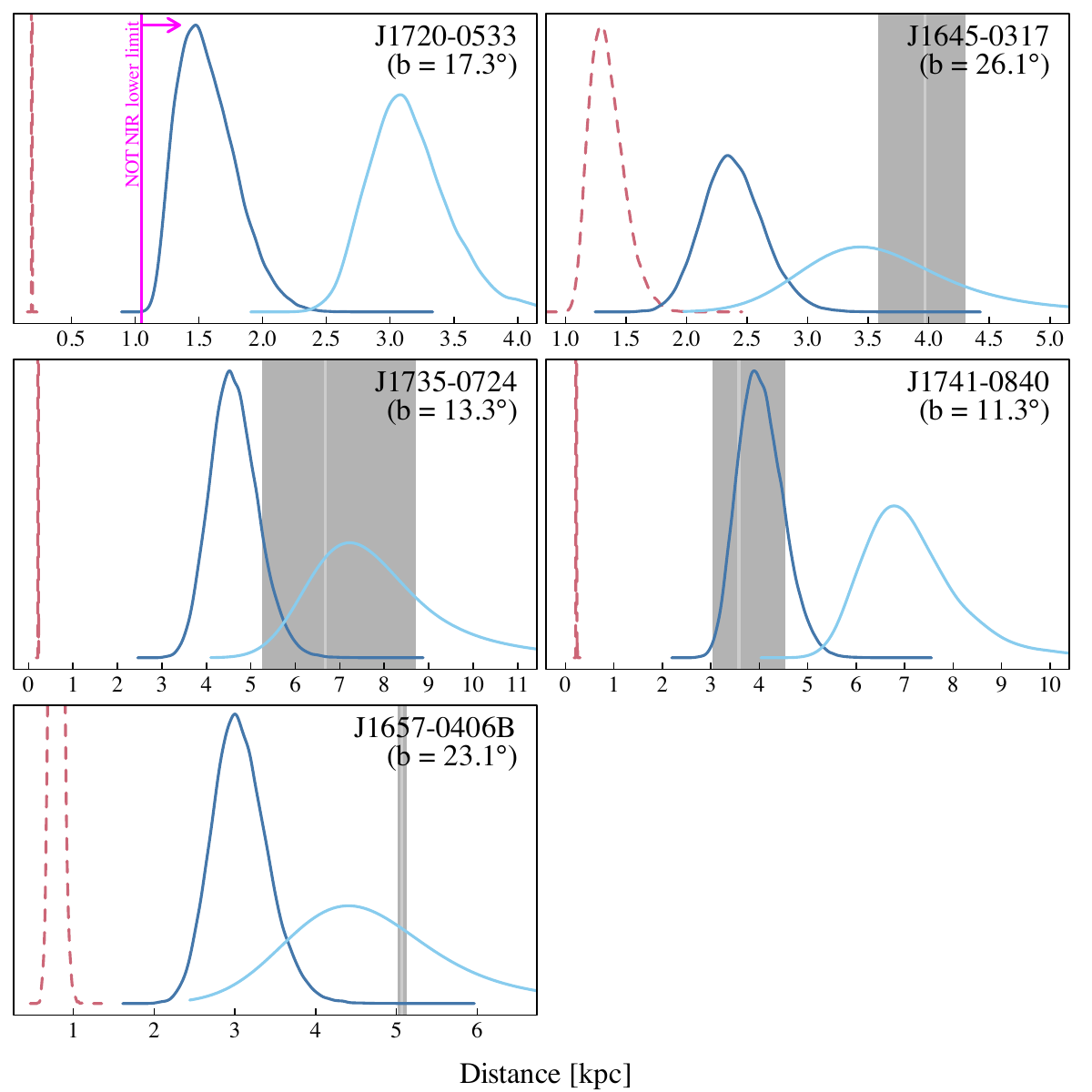}
    \caption{DM distance posteriors of recently discovered pulsars towards the NPS/Loop I. The plotting scheme is similar to Fig. \ref{fig:pulsars}. For PSR J1720$-$0534, we denote our lower limit on the distance based on the infrared non-detection with a vertical magenta line. Estimating the DM distances using only the Galactic thick disc component with updated parameters enhances the accuracy of DM distance estimation for PSR J1645$-$0317, PSR J1735$-$0724, and PSR J1657$-$0408B, and places PSR J1720$-$0534 at a distance of 3.1 kpc.}
    \label{fig:pulsars_new}
\end{figure}

\begin{table*}
    \centering
    \caption{Pulsars, their distance estimates, and line-of-sight properties in the direction of the NPS/Loop I ($5^{\circ}\lesssim l \lesssim 41^{\circ}$, $5^{\circ}\lesssim b \lesssim 38^{\circ}$). The columns display the source name, Galactic coordinates, dispersion measure (DM), YMW16 DM distance, YMW16 DM distance without the NPS/Loop I component, DM distance using only the Galactic thick disc component, NE2001 DM distance, independent distance measurement, average electron density in the line of sight based on this distance, the method of the independent distance measurement ($\Pi$ -- parallax, GC -- globular cluster (parallax), IR -- companion non-detection, H \textsc{I} -- kinematic distance), and its reference. The five sources at the upper part of the table were recently discovered and were not included in deriving the YMW16 electron density model.}\label{tab:pulsar_table}
    \begin{tabular}{lccccccccccc}
    \\[-1.8ex]\hline
    \hline \\[-1.8ex]
        Pulsar & $l$$^{a}$ & $b$$^{a}$ & DM$^{a}$ & $d_{\mathrm{YMW16}}$$^{b}$ & $d_{\mathrm{YMW16-NPS}}$$^{b}$ & $d_{\mathrm{disc}}$$^{b}$ & $d_{\mathrm{NE2001}}$$^{c}$ & $d_{\mathrm{other}}$ & $\bar{n}_{e}$ & Method & Ref$^{d}$ \\
        Name & (deg) & (deg) & (cm$^{-3}$ pc) & (kpc) & (kpc) & (kpc) & (kpc) & (kpc) & ($10^{-3}$ cm$^{-3}$) \\[1.0ex] 
        \hline\\[-1.8ex]
        J1645-0317 & 14.114 & 26.062 & 35.7 & 1.29$^{+0.14}_{-0.06}$ & 2.4$^{+0.3}_{-0.1}$ & 3.4$^{+0.6}_{-0.2}$ & 1.1 & 4.0$^{+0.3}_{-0.4}$ & 8.9 & $\Pi$ & 1 \\
        J1657-0406B & 15.137 & 23.076 & 43.4 & 0.77$^{+0.06}_{-0.03}$ & 3.0$^{+0.3}_{-0.1}$ & 4.4$^{+1.0}_{-0.4}$ & 1.7 & 5.07$^{+0.06}_{-0.06}$ & 8.6 & GC & 2 \\        
        J1720-0534 & 17.067 & 17.252 & 36.8 & 0.191$^{+0.001}_{-0.001}$ & 1.5$^{+0.2}_{-0.1}$ & 3.1$^{+0.3}_{-0.1}$ & 1.3 & $>$1.05 & $<$35.0 & IR & 3 \\
        J1735-0724 & 17.271 & 13.284 & 73.5 & 0.213$^{+0.002}_{-0.001}$ & 4.6$^{+0.5}_{-0.2}$ & 7.2$^{+1.4}_{-0.5}$ & 2.3 & 6.7$^{+2.0}_{-1.4}$ & 11.0 & $\Pi$ & 1 \\
        J1741-0840 & 16.955 & 11.304 & 74.9 & 0.222$^{+0.003}_{-0.001}$ & 3.9$^{+0.5}_{-0.2}$ & 6.8$^{+0.9}_{-0.4}$ & 2.2 & 3.6$^{+1.0}_{-0.6}$ & 20.8 & $\Pi$ & 1 \\
        \hline
        J1640+2224 & 41.051 & 38.271 & 18.43 & 1.49$^{+0.08}_{-0.03}$ & 1.50$^{+0.09}_{-0.04}$ & 1.59$^{+0.16}_{-0.07}$ & 1.16 & 1.4$^{+0.1}_{-0.1}$ & 7.7 & $\Pi$ & 4 \\
        J1643-1224 & 5.669 & 21.218 & 62.3 & 0.78$^{+0.07}_{-0.03}$ & 5.7$^{+1.1}_{-0.4}$ & --$^{e}$ & 2.4 & 0.95$^{+0.15}_{-0.11}$ & 84.3 & $\Pi$ & 4 \\
        J1713+0747 & 28.751 & 25.223 & 15.99 & 0.92$^{+0.04}_{-0.02}$ & 0.92$^{+0.04}_{-0.02}$ & 1.29$^{+0.11}_{-0.04}$ & 0.92 & 1.05$^{+0.06}_{-0.07}$ & 13.6 & $\Pi$ & 5\\
        J1721-1936 & 4.857 & 9.738 & 75.7 & 3.0$^{+0.3}_{-0.1}$ & 3.1$^{+0.3}_{-0.1}$ & 6.8$^{+0.9}_{-0.4}$ & 1.9 & 8.0$^{+0.2}_{-0.2}$ & 9.5 & GC & 2 \\
        J1738+0333 & 27.721 & 17.742 & 33.8 & 1.45$^{+0.17}_{-0.07}$ & 1.45$^{+0.17}_{-0.07}$ & 2.8$^{+0.3}_{-0.1}$ & 1.43 & 1.74$^{+0.15}_{-0.13}$ & 23.0 & $\Pi$ & 4 \\        
        J1741+1351 & 37.885 & 21.641 & 24.21 & 1.35$^{+0.09}_{-0.04}$ & 1.35$^{+0.09}_{-0.04}$ & 1.98$^{+0.18}_{-0.08}$ & 0.90 & 1.8$^{+0.5}_{-0.3}$ & 22.4 & $\Pi$ & 6 \\
        J1744-1134 & 14.794 & 9.180 & 3.14 & 0.148$^{+0.001}_{-0.001}$ & 0.221$^{+0.011}_{-0.004}$ & 0.270$^{+0.020}_{-0.008}$ & 0.41 & 0.44$^{+0.02}_{-0.02}$ & 7.9 & $\Pi$ & 6 \\
        J1801-0857A & 19.225 & 6.762 & 182.56 & 6.5$^{+0.7}_{-0.3}$ & 13.9$^{+1.6}_{-0.7}$ & --$^{e}$ & 4.8 & 9.2$^{+0.6}_{-0.5}$ & 19.8 & GC & 2 \\
        J1804-0735 & 20.792 & 6.773 & 186.32 & 8.0$^{+0.8}_{-0.3}$ & 14.8$^{+1.8}_{-0.7}$ & --$^{e}$ & 5.0 & 8.2$^{+0.4}_{-0.4}$ & 22.7 & GC & 2 \\
        J1807-0847 & 20.061 & 5.587 & 112.38 & 2.72$^{+0.20}_{-0.08}$ & 3.8$^{+0.3}_{-0.1}$ & 9.2$^{+0.9}_{-0.4}$ & 2.73 & 1.5$^{+1.2}_{-0.9}$ & 74.9 & H I & 7 \\
        J1820-0427 & 25.456 & 4.733 & 84.44 & 2.96$^{+0.20}_{-0.08}$ & 3.15$^{+0.17}_{-0.07}$ & 6.5$^{+0.6}_{-0.2}$ & 1.94 & 0.3$^{+0.6}_{-0.2}$ & 281.5 & H I & 7 \\
        J1823+0550 & 34.987 & 8.859 & 66.78 & 3.1$^{+0.4}_{-0.2}$ & 3.1$^{+0.4}_{-0.2}$ & 5.5$^{+0.5}_{-0.2}$ & 1.8 & 2.0$^{+1.3}_{-0.8}$ & 33.4 & H I & 7 \\
        \hline
        \multicolumn{11}{p{0.9\linewidth}}{\textit{$^{a}$} We collected the up-to-date Galactic coordinates and DMs from the Australia Telescope National Facility (ATNF \textsc{psrcat v1.70}; \url{https://www.atnf.csiro.au/research/pulsar/psrcat}).}\\
        \multicolumn{11}{p{0.9\linewidth}}{\textit{$^{b}$} We calculated the distance estimates from the DM using \textsc{PSRdist} \citep{bartels18}.}\\
        \multicolumn{11}{p{0.9\linewidth}}{\textit{$^{c}$} We calculated the distance estimates from the DM using \textsc{PyGEDM} \citep{price21}.}\\
        \multicolumn{11}{p{0.9\linewidth}}{\textit{$^{d}$} \textbf{References:} 1) \citet{deller19}, 2) \citet{baumgardt21} 3) This work, 4) \citet{ding23}, 5) \citet{chatterjee09} 6) \citet{arzoumanian18}, 7) \citet{frail90}.}\\
        \multicolumn{11}{p{0.9\linewidth}}{\textit{$^{e}$} The DM distance is not constrained.}
    \end{tabular}
    \label{tab:pulsars}
\end{table*}

\begin{figure*}
    \centering
    \includegraphics[width = 1.6\columnwidth]{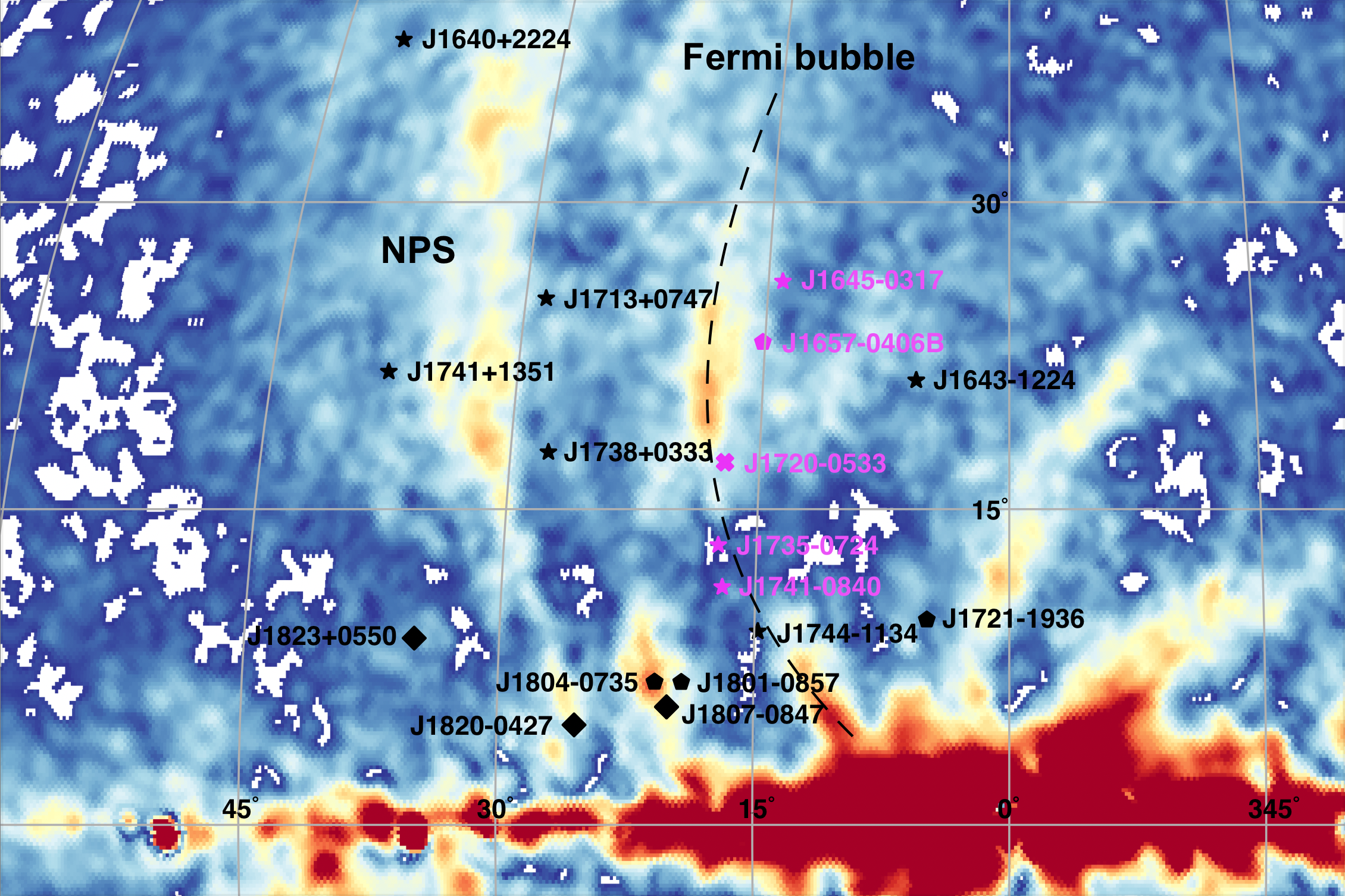}
    \includegraphics[width = 1.6\columnwidth]{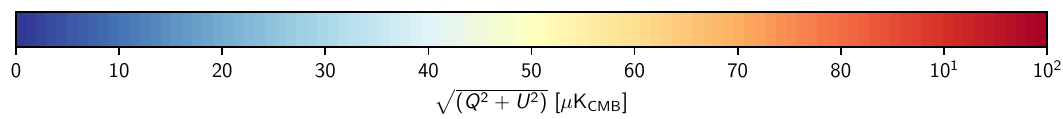}    
    \includegraphics[width = 1.6\columnwidth]{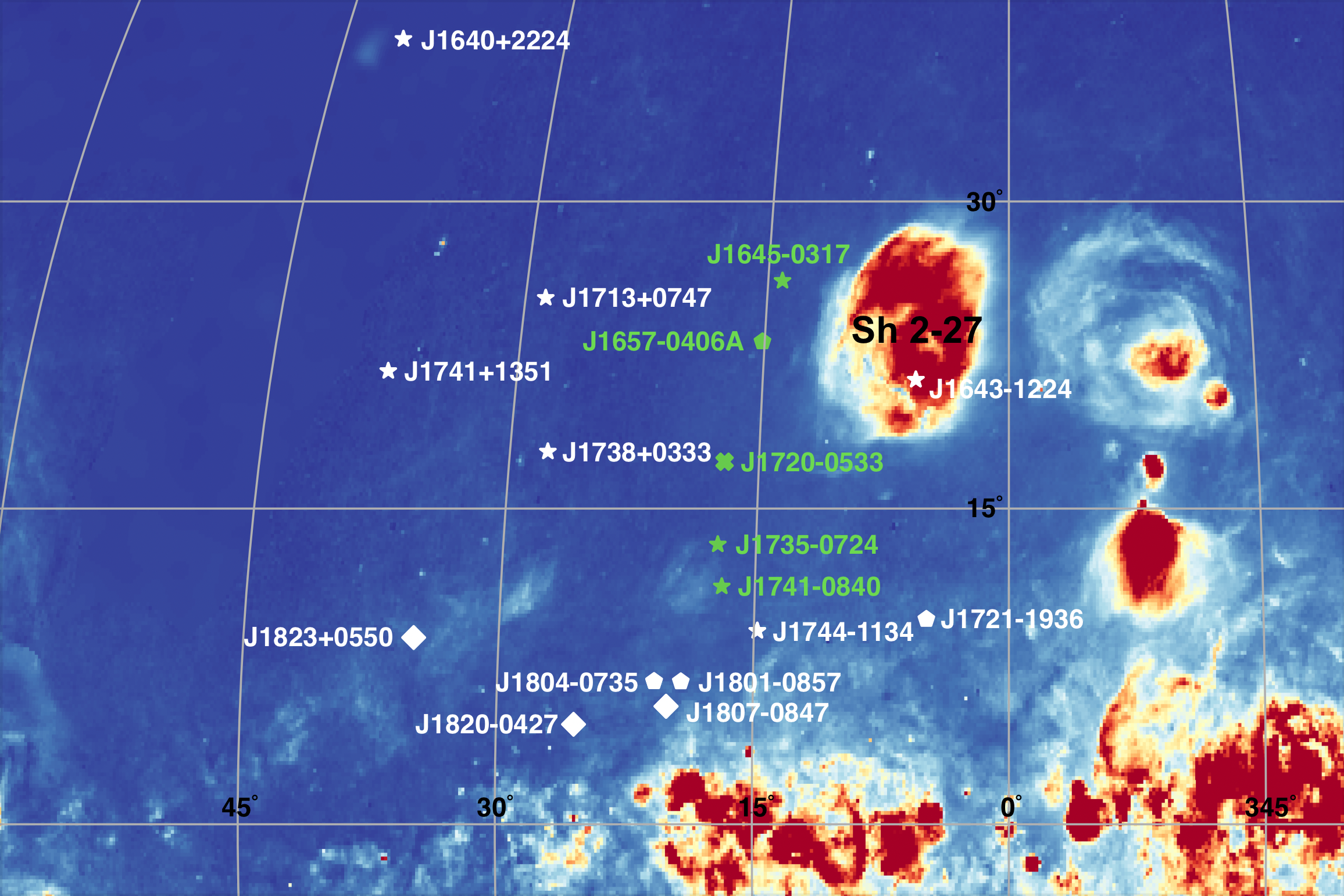}
    \includegraphics[width = 1.6\columnwidth]{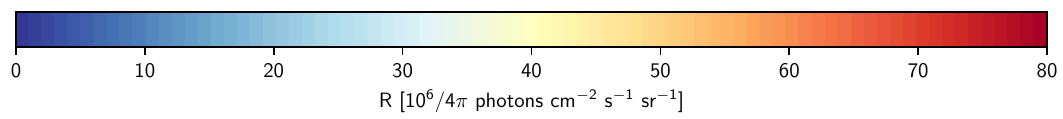}    
    \caption{Pulsar locations in the direction of NPS/Loop I plotted on the \textit{Planck} polarization intensity map at 30 GHz \citep[][\textit{top panel}]{planck16} and the H$\alpha$ map \citep[][\textit{bottom panel}]{finkbeiner03}. The coordinate grid is in Galactic coordinates, with labels marked along the coordinate lines in the figures. Different symbols denote the distance measurement method (star - parallax, pentagon - globular cluster, diamond - H \textsc{I}  absorption), and the location of PSR J1720$-$0534 is marked with a cross. The symbol colours indicate their inclusion in the electron density model (black/white: included in YMW16, magenta/green: not included in YMW16). The northern Fermi bubble (black dashed lines) and NPS region are indicated in the \textit{Planck} map, and the H \textsc{II} region Sh 2-27 is denoted in the H$\alpha$ map.}
    \label{fig:pulsar_map}
\end{figure*}

The NPS/Loop I component was included by YMW16 to improve the model fit in the region of the sky roughly encompassing the area $0^{\circ}\lesssim l, b \lesssim 30^{\circ}$ that contained pulsars with overestimated distances. \citet{yao17} modeled this component with a hemispherical cap centered 195 pc from the Sun. However, only 11 pulsars drive the fit in this direction. Their best-fit solution resulted in eight pulsars having a good match with independent distance measurements, and three pulsars with under- or overpredicted DM distances. Only 1 of the 11 pulsars (PSR J1643$-$1224) is located at higher latitudes ($b>10^{\circ}$), where more than half of the dispersing component lies. Furthermore, out of the eight pulsars with matching distances, four are situated in the Galactic plane ($-1^{\circ}\lesssim b \lesssim 1^{\circ}$), and three are in close proximity to it ($b\sim6^{\circ}$). 

Consequently, PSR J1643$-$1224, positioned at higher latitudes, likely influences the overall shape chosen for the entire component. Only the electron density of this large hemispherical cap representing the NPS/Loop I was allowed to vary in the model fits of YMW16, resulting in a high value (1.9 cm$^{-3}$; see their Table 2 and sections 3.7 and 5.1). Also, it is worth noting that the location of this component (restricted to $0^{\circ}\lesssim l, b \lesssim 30^{\circ}$) does not coincide with the brightest part of NPS/Loop I ($l > 30^{\circ}$, $b > 15^{\circ}$) that supposedly would be the densest region. Instead, its position is more consistent with the Fermi Bubbles. 

Thus, we conclude that the large and dense NPS/Loop I component in the YMW16 model can severely underestimate the distances to some pulsars in that direction. We find that in the case of PSR J1720$-$0534, YMW16 underestimates its distance at least by a factor of 5.

We tested the necessity of the NPS/Loop I component in YMW16 by excluding it from the model and re-evaluating the DM distances for PSR J1720$-$0534 and a sample of pulsars in the direction of NPS/Loop I (see Table \ref{tab:pulsar_table}). Recently discovered pulsars are listed in the upper part of the table, while those included in YMW16 are listed in the lower part. We limited our analysis to sources with Galactic latitudes $b\gtrsim5^{\circ}$ to avoid complications arising from features near the Galactic plane, such as an increasing number of clumps and voids in the local interstellar medium along the line of sight. 

Figs \ref{fig:pulsars} and \ref{fig:pulsars_new} display the distance posteriors of pulsar DM distances using the YMW16 model, both with and without the NPS/Loop I component (depicted as dashed red and solid dark blue curves, respectively). Independent distance measurements are represented by solid vertical lines, with 1-$\sigma$ errors indicated by grey bands in Figs \ref{fig:pulsars} and \ref{fig:pulsars_new}, and tabulated in Table \ref{tab:pulsar_table}.

Pulsars PSR J1804$-$0735, PSR J1807$-$0847, and PSR J1643$-$1224 (Fig. \ref{fig:pulsars}) exhibit a need for increased electron density modelled with the NPS/Loop I component to align with independent distance measurements. While the former two are in proximity to the Galactic plane ($b\sim6^{\circ}$) and may experience additional dispersion from the Galactic thin disc, spiral arms, and/or fluctuations in line-of-sight electron densities, PSR J1643$-$1224 is situated at much higher Galactic latitudes ($b=21^{\circ}$). However, this region of the sky hosts a robust H \textsc{II} region \citep[Sharpless 2-27 -- Sh 2-27; ][see also Fig. \ref{fig:pulsar_map}]{gvaramadze12}, originating from the $\zeta$-Oph O9.5 V star positioned at a distance of 112 pc \citep{vanleeuwen07}. This H \textsc{II} region has an almost circular shape with a radius of 5$^{\circ}$ or 10 pc at the star's distance, and an average electron number density of $\sim$3 cm$^{-3}$ \citep{gull79}. Consequently, the additional density observed towards PSR J1643$-$1224 can be attributed to this intervening H \textsc{II} region \citep{mall22,ocker20}, obviating the need to invoke NPS/Loop I.

Considering the more recently discovered pulsars, excluding the NPS/Loop I component brings the resulting DM distances closer to the independent distance measurements in all cases (Fig. \ref{fig:pulsars_new}). For PSR J1735$-$0724 and PSR J1741$-$0840, the DM distance posteriors overlap with the parallax distances. However, for other sources, the distances are still underestimated. In addition, the DM distance for PSR J1720$-$0534 changes from 0.19 kpc to 1.5 kpc.

Based on our analysis of pulsar distances towards the NPS/Loop I component in YMW16, we conclude that the removal of this component overall improves the DM distance estimates in this region of the sky, especially for sources not too close to the Galactic plane. However, there are still discrepancies, and many pulsars still have underpredicted DM distances compared to parallax distances. This suggests that the free electron densities in the YMW16 model towards the NPS/Loop I region are still in many cases too high.

Since the publication of the YMW16 electron density model, there has been an update on the Galactic thick disc parameters (density and scale height) by \citet{ocker20} that differ from the values used in YMW16. We note, however, that the functions used to model the disc differ slightly between \citet{ocker20} and YMW16, where the former uses an exponential, and the latter uses a hyperbolic secant function. Using values closely matching the exponential function in \citet{ocker20}, we update the corresponding mid-plane electron density using the hyperbolic secant function to $n_{0}=0.013\pm0.001$ cm$^{-3}$ (YMW16 used $n_{0}=0.0113\pm0.0004$ cm$^{-3}$) while the scale height remains the same as for YMW16: $z_{0}=1.67\pm0.15$ kpc. Using only the Galactic thick disc component in the electron dispersion model places the DM distances to pulsars at the high Galactic latitudes ($b\gtrsim10^{\circ}$) more in line with the independent distance estimates (depicted by solid light blue lines in Figs \ref{fig:pulsars} and \ref{fig:pulsars_new}), with the exception of PSR J1643$-$1224, PSR J1738$+$0333, and PSR J1741$-$0840. PSR J1738$+$0333 is located close to the brightest radio polarization region in NPS (see Fig. \ref{fig:pulsar_map}) and could imply an intervening dense region towards the brightest component of NPS/Loop I. PSR J1741$-$0840, having relatively low Galactic latitude ($b=11.3^{\circ}$), is likely affected by other Galactic components, as exemplified by the matching YMW16 DM distance with the NPS/Loop I component removed (Fig. \ref{fig:pulsars_new}). Assuming just the thick disc component in the electron dispersion model moves the DM distance of PSR J1720$-$0534 to 3.1 kpc.

\subsection{Implications for the distance to PSR J1720$-$0534}\label{sec:dist_to_J1720}

Taking the YMW16 distance at face value would demand very exceptional properties for both the companion and the emission mechanisms of the neutron star in PSR J1720$-$0534, considering our upper limits, especially in the infrared and $\gamma$-rays (Section 3, Fig. \ref{fig:sed}). Based on radio-timing observations, all derived properties are typical of BW CBMPs \citep[][e.g., $\dot{E}=9.2\times10^{34}$ erg/s, $P_{\rm orb}=3.16$ hr, $M_{\rm c,min}=0.03 \, M_{\odot}$]{miao23}.

Considering first the companion star, the average temperature over the orbit would need to be less than 1400 K to accommodate the infrared non-detection and the YMW16 distance, which is already 900 K less than the lowest known average temperature of a BW companion in the literature \citep{matasanchez23}. Given the short orbital period and fairly average spin-down power, irradiation of the companion star by the pulsar wind and heating of the stellar surface facing the pulsar is expected \citep{turchetta23}. To estimate the strength of irradiation, we can use the pulsar spin-down to companion flux ratio at the location of the companion star \citep{turchetta23}:

\begin{equation}
f_{\rm sd} \equiv \frac{\dot{E}}{L_{2}}\frac{R_{2}^2}{a^2} \simeq 7700 \, \dot{E}_{34} \, T_{\rm b,1000}^{-4} \, P_{\rm orb, hr}^{-4/3} \simeq 1528 \, T_{\rm b,1000}^{-4},
\end{equation}

\noindent where we have used the above values for the spin-down power and orbital period of PSR J1720$-$0534. Thus, for the low companion effective base temperatures required by the close distance ($T_{\rm b} \sim 1000-2000$ K), the companion star would exhibit strong irradiation ($f_{\rm sd}\gtrsim$100) and subsequently a much higher average temperature, resulting in higher luminosities that would be observable. In this case, we would also detect the bright/irradiated side of the companion in at least some of the near-infrared images, but we do not (Section \ref{sec:nir}).

Secondly, considering the YMW16 distance with the $\gamma$-ray flux upper limit would imply a $\gamma$-ray efficiency well below 1 per cent, which is unprecedented in the millisecond pulsar population. This would require an intrinsic spin-down power lower than the $\gamma$-ray death line \citep[$\dot{E}<10^{33}$ erg/s;][]{kalapotharakos18} and subsequently high transverse proper motion to lower the measured spin-down power below this value through Shklovskii correction. On the other hand, an unfavourable viewing angle, which places our line of sight outside the $\gamma$-ray beam of the pulsar, could result in $\gamma$-ray non-detection as well.

Therefore, we conclude that instead of being a very exceptional BW, the non-detections from PSR J1720$-$0534 can be understood by placing the source at a much larger distance than what the YMW16 model implies. This requires the modification of the electron density model of YMW16 by removing the NPS/Loop I component and updating the Galactic thick disc parameters. Making these changes in the YMW16 model also aligns with the comparison of parallax distances to DM distances in other nearby pulsars to PSR J1720$-$0534 (taking into account the caveats mentioned in Section 4.1 about intervening H \textsc{II} regions and sources close to the Galactic plane).

Given a likely distance of 3.1 kpc to PSR J1720$-$0534, it is possible to accommodate a Roche lobe filling stellar companion ($R\sim0.16\,M_{\odot}$) with an average temperature of 2500 K below the infrared limiting magnitude (see Fig. \ref{fig:sed}). On the other hand, if the radius of the companion is smaller ($R\lesssim0.09\,M_{\odot}$), subsequently the average temperature could be higher, $\sim$4000 K, that is close to the global average \citep{matasanchez23}. Nevertheless, the likely $J$-band magnitude of the companion should be on the order of $\sim$24--25 mag, which would be still observable with $\sim$2-m size telescopes. However, in the optical, due to higher extinction and spectral curvature, the estimated magnitudes are higher, $\gtrsim$25 mag, depending strongly on the used filter and the temperature of the companion star. For X-rays, typical BW X-ray luminosities of 10$^{30}$--10$^{31}$ erg s$^{-1}$ correspond to fluxes [0.6--6]$\times10^{-15}$ erg/s/cm$^2$ for a distance of 3.1 kpc and using the hydrogen column density from Section 3.3, which is detectable, e.g., with \textit{XMM-Newton}.  

\section{Conclusions}

In this paper, we conducted an extensive multiwavelength analysis of the region around PSR J1720$-$0534, a CBMP located in the direction of the NPS. Despite thorough investigations using 14 yr of \textit{Fermi}/LAT data, 2.7-h optical and 3.5-h near-infrared observations with the LCO and the NOT, and \textit{Swift}/XRT pointing observations, no significant counterparts were detected in gamma-ray, optical, near-infrared, or X-ray wavelengths.

Our near-infrared observations provided a deep upper limit on the magnitude of the potential counterpart, leading to a conservative lower limit for the distance of PSR J1720$-$0534: $d>$ 1.05 kpc. This constraint challenges the close distance estimate from the DM using the electron density model of YMW16.

Furthermore, our analysis of pulsar distances indicates that the inclusion of a dense component in the electron density distribution towards the NPS in the YMW16 model likely contributes to this discrepancy. Removing this component improves the accuracy of DM distance estimates in this region, especially for sources not too close to the Galactic plane. However, discrepancies still persist, indicating that the electron density distribution in this region is still overestimated in the model likely arising from the parameters of the thick disc and other Galactic components such as spiral arms, the Galactic thin disc, and complex small-scale structures in the line of sight close to the Galactic plane. We suggest that the electron distribution models in the future should include prominent H \textsc{II} regions, possibly reinstate small-scale structure in the local interstellar medium, and update the parameters for the large-scale Galactic components. Our study highlights the importance of refining electron density models for a more accurate understanding of pulsar distances in the Galaxy.

\section*{Acknowledgements}
The authors would like to thank Michael Unger for discussions on H$\alpha$ maps and their connection to dispersion measure. We also thank J.B. Wang and C.C. Miao for sharing the pulsar location ahead of publication. This project has received funding from the European Research Council (ERC) under the European Union’s Horizon 2020 research and innovation programme (grant agreement No. 101002352, PI: M. Linares). The observations were conducted using the Nordic Optical Telescope, owned in collaboration by the University of Turku and Aarhus University, and operated jointly by Aarhus University, the University of Turku and the University of Oslo, representing Denmark, Finland and Norway, the University of Iceland and Stockholm University at the Observatorio del Roque de los Muchachos, La Palma, Spain, of the Instituto de Astrofisica de Canarias. This work also makes use of observations from the Las Cumbres Observatory global telescope network, and is based on observations obtained with Planck (\url{http://www.esa.int/Planck}), an ESA science mission with instruments and contributions directly funded by ESA Member States, NASA, and Canada. This research has made use of data and/or software provided by the High Energy Astrophysics Science Archive Research Center (HEASARC), which is a service of the Astrophysics Science Division at NASA/GSFC. In particular, we acknowledge the use of public data from the Swift data archive. We gratefully acknowledge the use of \textsc{PSRdist}, available at \url{https://github.com/tedwards2412/PSRdist} and \textsc{PyGEDM}, available at \url{https://apps.datacentral.org.au/pygedm}. 
\section*{Data Availability}
The data underlying this article are available in the \textit{Fermi}-LAT data server at \url{https://fermi.gsfc.nasa.gov/ssc/data/access/}. \textit{Fermi}-LAT analysis results as well as the NOT near-infrared and LCO optical data can be shared upon reasonable request from the authors. The \textit{Swift} X-ray data are available at HEASARC (\url{https://heasarc.gsfc.nasa.gov}). The \textit{Planck} polarization map is available at the Planck Legacy Archive (\url{http://pla.esac.esa.int/pla}). The H$\alpha$ map is available at \url{https://faun.rc.fas.harvard.edu/dfink/skymaps/halpha}.



\bibliographystyle{mnras}
\bibliography{bibliography} 




\appendix

\section{The extended Fermi-LAT source FHES J1723.5$-$0501}\label{sec:extended source}
\subsection{Data analysis}

\begin{table}
    \centering
    \caption{Summary of the \textit{Fermi}-LAT data selections and analysis configuration.}
    \begin{tabular}{ll}
    \\[-1.8ex]\hline
    \hline \\[-1.8ex]
        Selection & Criterion \\[1.0ex] 
        \hline\\[-1.8ex]
        Observation period & August 4, 2008, to September 26, 2022 \\
        Mission Elapsed Time (MET) & 239557417 to 685859961 \\
        Central coordinates & $l=17.90^\circ$, $b = 16.96^\circ$ \\
        Radius & $8^\circ$ \\
        Energy range & $1-1000$\,GeV \\
        Zenith angle & $z \leq 100^\circ$ \\
        Event types & $ \texttt{evtype}=32 $ and $\texttt{evtype} = 28 $\\
        Event class & $\texttt{evclass}$ = 128\\
        Data quality cut & DATA\_QUAL == 1 \\ 
        & LAT\_CONFIG == 1 \\[1.0ex] \hline
    \end{tabular}
    \label{tab:dataselec}
\end{table}

We model the extended \textit{Fermi}-LAT source using a ROI optimization algorithm based on the algorithm presented in \citet{ackermann18}, where the authors presented the first \textit{Fermi} High-Latitude Extended Sources Catalog (FHES) and reported the discovery of 19 new extended sources, including FHES J1723.5$-$0501 (4FGL J1723.5$-$0501e). The data selection and configuration used to analyse the ROI are explained in Section \ref{sec:obsanddata} and summarized in Table \ref{tab:dataselec}. The analysis starts from a baseline source model with the Galactic and isotropic background models, together with the 4FGL catalog sources within a $10^\circ \times 10^\circ$ region centered at the position of FHES J1723.5$-$0501. The extended source is then removed from the source model. Next, we change the spectral model of all catalog sources with $\mathrm{TS}>100$ modelled with the power law (PL) spectral parameterization to a log-parabola (LP). We do this to ensure accurate modelling of background sources with undetected spectral curvature. As PL is a special case of LP ($\beta=0$), this comes without loss of generality. 

Once all baseline model sources are configured, we perform a spectral fit of the flux normalization and spectral shape of the Galactic diffuse emission model, and all point sources with at least one predicted photon ($n_\mathrm{pred}\geq1$) from the catalog parameters. Next, we relocalize the sources inside the ROI with a distance of at least $0.1^\circ$ from the ROI boundary to their local TS peak and refit their normalizations simultaneously. Finally, we refit the spectral parameters of all model components inside the ROI to complete the optimization of the baseline model.

The analysis then proceeds iteratively by looking for new candidate point sources. First, we investigate sources in the outer ROI defined by $R>R_\mathrm{inner}$, where $R_\mathrm{inner} = 1^\circ$. Candidates are identified by creating a TS map for a test source with a PL spectral model with index $\Gamma=2$. Starting from the peak with the highest TS value, we add candidate point sources with $\mathrm{TS}>9$ to the model, as long as the new candidate is at least $0^\circ5$ away from an existing candidate source with a higher TS value. To ensure that the source is bright enough to detect the spectral curvature parameters in the LP case, we only model candidate sources with a LP spectral shape for sources with $\mathrm{TS}>100$, and otherwise they are modeled with a PL. When the candidate sources are added to the model, we simultaneously fit their spectral shapes and normalizations. Once all candidate sources fulfilling our criteria are added to the model, a new TS map is generated and new candidate sources are added in the same way. This procedure continues until there are no candidate sources left located at $R>R_\mathrm{inner}$ with $\mathrm{TS}>9$. Finally, we refit the normalizations and spectral shapes of all model components to complete the optimization of the outer ROI. 

In the final part of the analysis in \citet{ackermann18}, they optimize the inner ROI by carefully looking for new point source candidates while they test the central source for extension. However, as there are no significant candidate point sources with $\mathrm{TS}<9$ within $R_\mathrm{inner}$ for this ROI, the final steps are not necessary.

\begin{figure}
    \centering
    \includegraphics[width=\columnwidth]{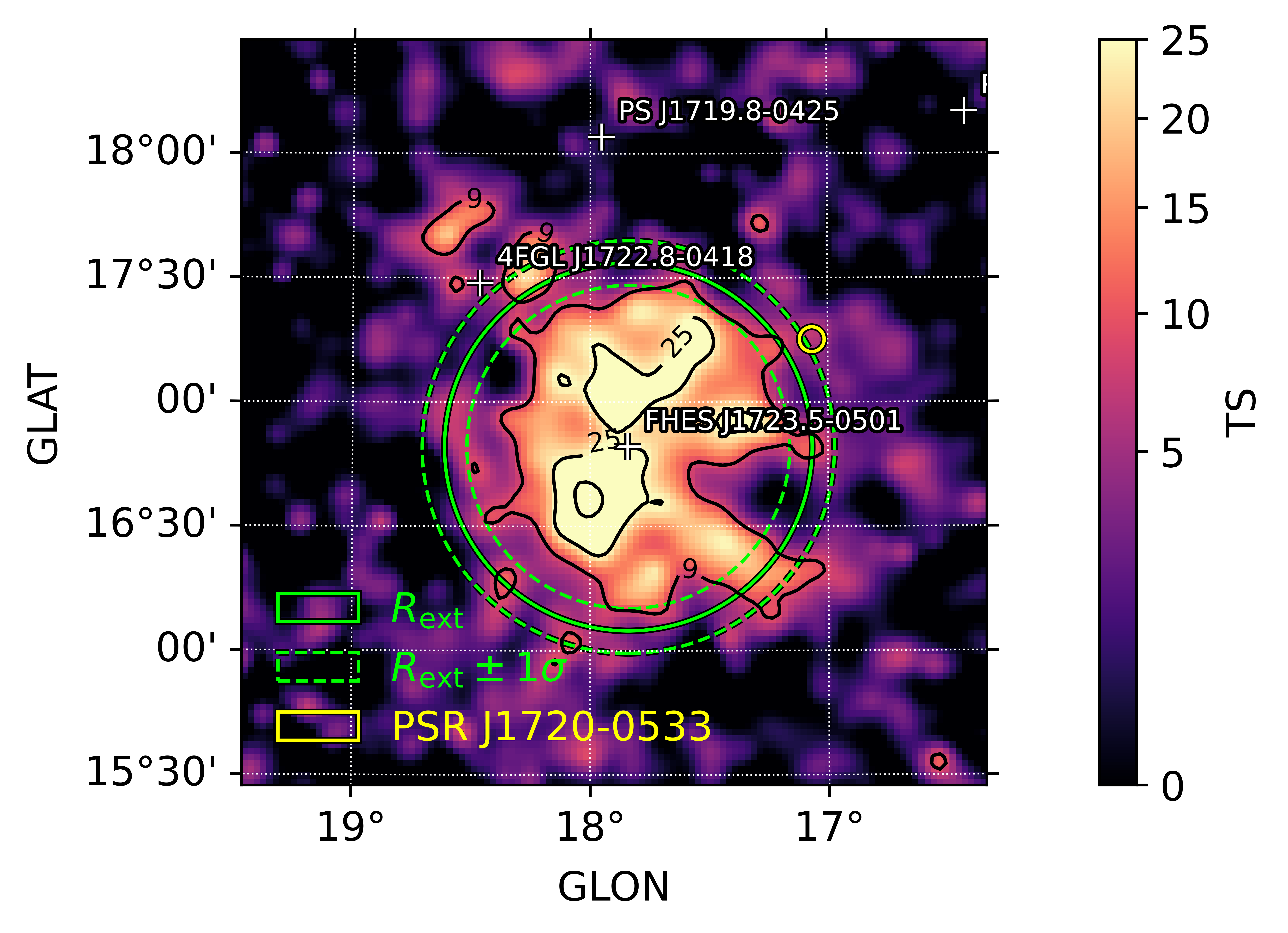}
    \caption{TS map of FHES J1723.5$-$0501 made by a test source with a power-law spectral model with $\Gamma=2$ and excluding the extended source from the source model. The green circles indicate the 68 per cent containment radius of the extension with $\pm1\sigma$ uncertainties. The yellow circle indicates the position of PSR J1720$-$0534, with a $\sim3$ arcmin uncertainty.}
    \label{fig:ext}
\end{figure}

The detection significance of extended emission is quantified by \begin{equation}
    \mathrm{TS}_\mathrm{ext}=2(\mathrm{ln}\mathcal{L}_{\mathrm{ext}+n}-\mathrm{ln}\mathcal{L}_{n}),
\end{equation}
which is the likelihood ratio between a model with an extended central component and a model with a central point source. \citet{ackermann18} classify the sources as extended if $\mathrm{TS}_\mathrm{ext}>16$, corresponding to a $4\sigma$ detection. For FHES J1723.5$-$0501, we obtain $\mathrm{TS}_\mathrm{ext}=115$. Therefore, we update our model with the extended central source by performing an extension fit. Then, we refit the normalization and spectral shape of all model components. Finally, we consider the ROI to be fully optimized by running a new extension fit of the central source and once again refit all spectral parameters of the model components. The extension fit finds, among other things, the best-fit position of the extension, together with the extension radius $R_\mathrm{ext}$ that is parameterized by the intrinsic 68 per cent containment radius of the source. The TS map of FHES J1723.5$-$0501 after the optimization of the ROI is presented in Figure \ref{fig:ext}.

\begin{table*}

\begin{tabular}{l l l l l l l l l}
\hline \hline
Analysis  & $l[^\circ]$ & $b [^\circ]$ & TS & $\mathrm{TS}_\mathrm{ext}$ & $R_{\mathrm{ext}}[^\circ]$ &$F_\gamma$\textsuperscript{a}& $G_\gamma$\textsuperscript{b}  & $\Gamma$ \\ \hline 
\citet{ackermann18} &  17.90   & 16.96    &  89.5  &  52.9   & $0.73 \pm 0.10 \pm 0.01$  &  $18.3 \pm 2.5 \pm 2.1  $            & - &  $ 1.97 \pm 0.08 \pm 0.06 $   \\ 
\citet{araya22} & -  & -  &  153.2  &  65.6   & $0.68^{+0.07}_{-0.16}$ &  -  & $\sim 1$&   $1.83 \pm 0.02 \pm 0.05$ \\ 
This work     &   17.84  &  16.82   &  133.8  & 114.7   &   $0.74^{+0.10}_{-0.08}$  & $16.2\pm2.0$ & $ 2.1 \pm 0.3  $   &  $1.93 \pm 0.07 $    \\ \hline
\end{tabular}%

    \begin{description}
    \small
         \item  $\textsuperscript{a}\gamma$-ray photon flux in units of $\SI{E-10}{cm^{-2}\,s^{-1}}$ in the $0.1-\SI{100}{GeV}$ energy band.
        \item $\textsuperscript{b}\gamma$-ray energy flux in units of $\SI{E-11}{erg\,cm^{-2}\,s^{-1}}$ in the $0.1-\SI{100}{GeV}$ energy band.
    \end{description}

\caption{The results from the extension fit of FHES J1723.5$-$0501 from \citet{ackermann18} and \citet{araya22}, compared to the results of the entire $2008-2022$ data set analysis of this work.}
\label{tab:fulldataext}

\end{table*}
\subsection{Discussion}
\citet{ackermann18} reported an unclassified $\SI{1.4}{GHz}$ radio shell engulfed by the extended emission, suggesting its association with a Type 1a SNR or a pulsar wind nebula (PWN). In a recent study, \citet{araya22} classified FHES J1723.5$-$0501 as a Type 1a SNR, naming it G17.8+16.7, and estimating its distance to be in the range of $d=1.4 - \SI{3.5}{kpc}$, using the characteristic $\SI{1.4}{GHz}$ radio luminosity range for SNRs and SNR evolutionary models. In their analysis of this radio emission, they calculate a two-point spectral index from the $1.4$ and $\SI{2.3}{GHz}$ radio flux densities of $\alpha=-0.75 \pm 0.15$ for $S\propto \nu^\alpha$, and conclude that this is consistent with non-thermal emission from a synchrotron-emitting shell SNR.

\begin{figure}
    \centering
    \includegraphics[width=\columnwidth]{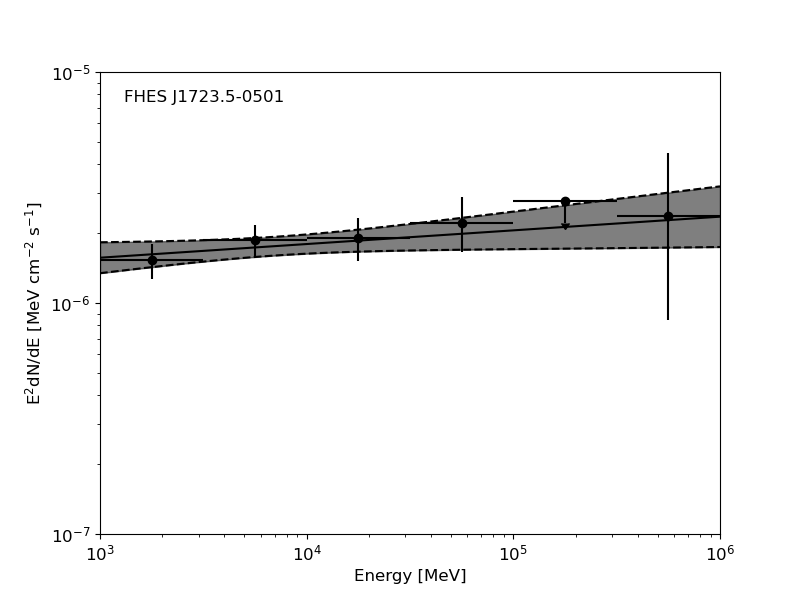}
    \caption{Spectral energy distribution for FHES J1723.5$-$0501 in the analysis energy range. The points indicate the measured $E^2\mathrm{d}N\mathrm{d}E$ values with uncertainties. The arrow shows an upper limit and the shaded area indicates the model spectrum with $\pm 1\sigma$ uncertainties.}
    \label{fig:SED}
\end{figure}

Our analysis results in a best-fit position of $l=17.84^\circ$ and $b=16.82^\circ$, with the extension radius of $R_\mathrm{ext}=0.74^\circ\pm0.09^\circ$ and $\mathrm{TS}=134$. We find a $\gamma$-ray photon flux of $F_\gamma=(16.2\pm2.0)\times\SI{E-10}{cm^{-2}\,s^{-1}}$, a $\gamma$-ray energy flux of $G_{\gamma}=(2.1\pm0.3)\times \SI{E-11}{erg\,cm^{-2}\,s^{-1}}$, and a PL spectral index of $\Gamma=1.93\pm0.07$. These values are fully consistent with the results from \citet{ackermann18} and \citet{araya22}. However, \citet{araya22} obtain a larger TS value compared to this work, but their $\mathrm{TS}_\mathrm{ext}$ value is only 57 per cent of ours. We suspect our larger $\mathrm{TS}_\mathrm{ext}$ value to originate from the use of a joint likelihood analysis with the PSF partition. The results from the extension fit of FHES J1723.5$-$0501 for this work compared to the two mentioned analyses are summarized in Table \ref{tab:fulldataext}. Figure \ref{fig:SED} shows the LAT spectrum of the extended source.

\begin{figure}
    \centering
    \includegraphics[width=\columnwidth]{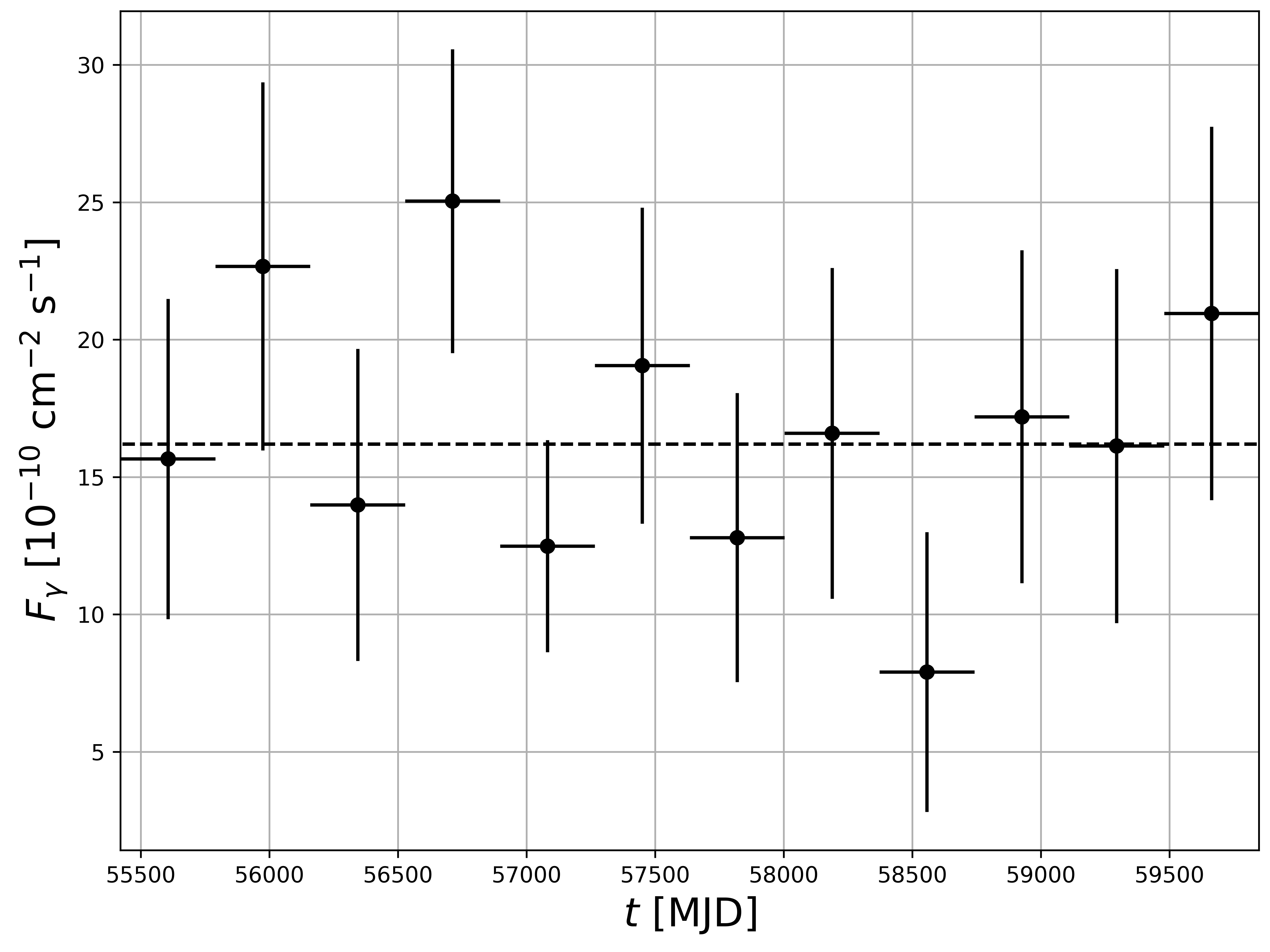}
    \caption{\textit{Fermi}-LAT lightcurve of FHES J1723.5$-$0501 between 2010 and 2022, produced using the optimized ROI model with the baseline analysis configurations. The dashed line indicates the average photon flux $F_\gamma=16.2\times10^{-10}$\,cm$^{-2}$\,s$^{-1}$. We observe no evident variability in the photon flux in the \textit{Fermi}-LAT data.}
    \label{fig:lc}
\end{figure}

Based on our observed $\gamma$-ray energy flux in the $0.1-\SI{100}{GeV}$ energy band and the distance estimate range of $d\simeq1.4-\SI{3.5}{kpc}$ from \citet{araya22}, we calculate the $\gamma$-ray luminosity of FHES J1723.5$-$0501 to be in the range of $L_\gamma= \SI{5E33}{}-\SI{3E34}{erg\,s^{-1}}$ that places it among the brightest SNR $\gamma$-ray luminosities observed, as can be seen in fig. 13 in the First \textit{Fermi}-LAT Supernova Remnant Catalog \citep[1SC;][]{acero16}. The calculated spectral index of $\Gamma=1.93\pm0.07$ places this potential $\gamma$-ray emitting SNR among the SNRs with the hardest reported spectral indices (see fig. 8 in 1SC). Finally, we extracted the long-term \textit{Fermi}-LAT light curve of this extended source and find it is consistent with a constant flux over 12 yr (Figure \ref{fig:lc}).

\section{Near-infrared study of the nearby source to PSR J1720-0534}\label{sec:varIR}

\begin{figure}
    \centering
    \includegraphics[width=\columnwidth]{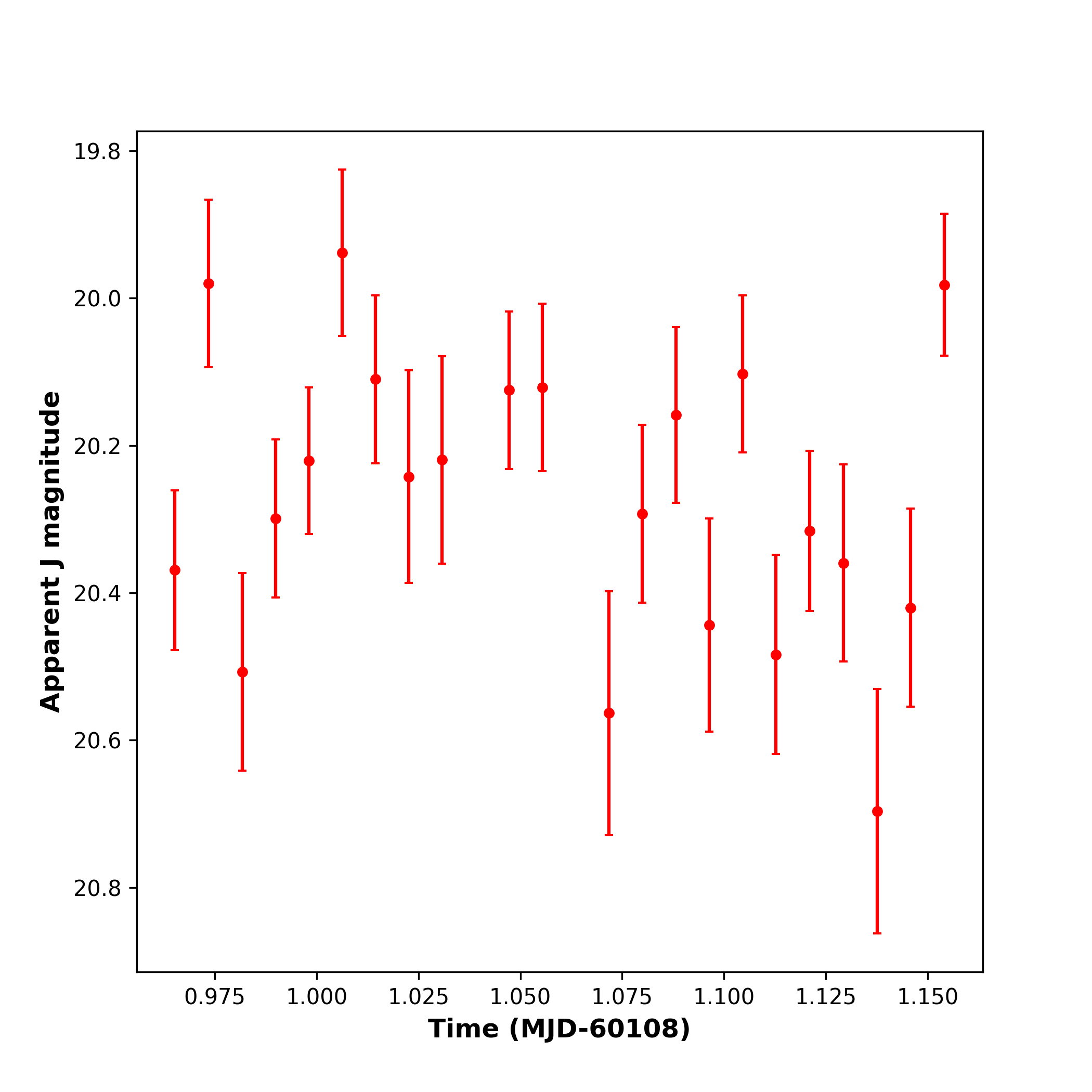}
    \caption{The $J$-band light curve of the nearest infrared source to the radio location of PSR J1720$-$0534.}
    \label{fig:IRlc}
\end{figure}

\begin{figure}
    \centering
    \includegraphics[width=\columnwidth]{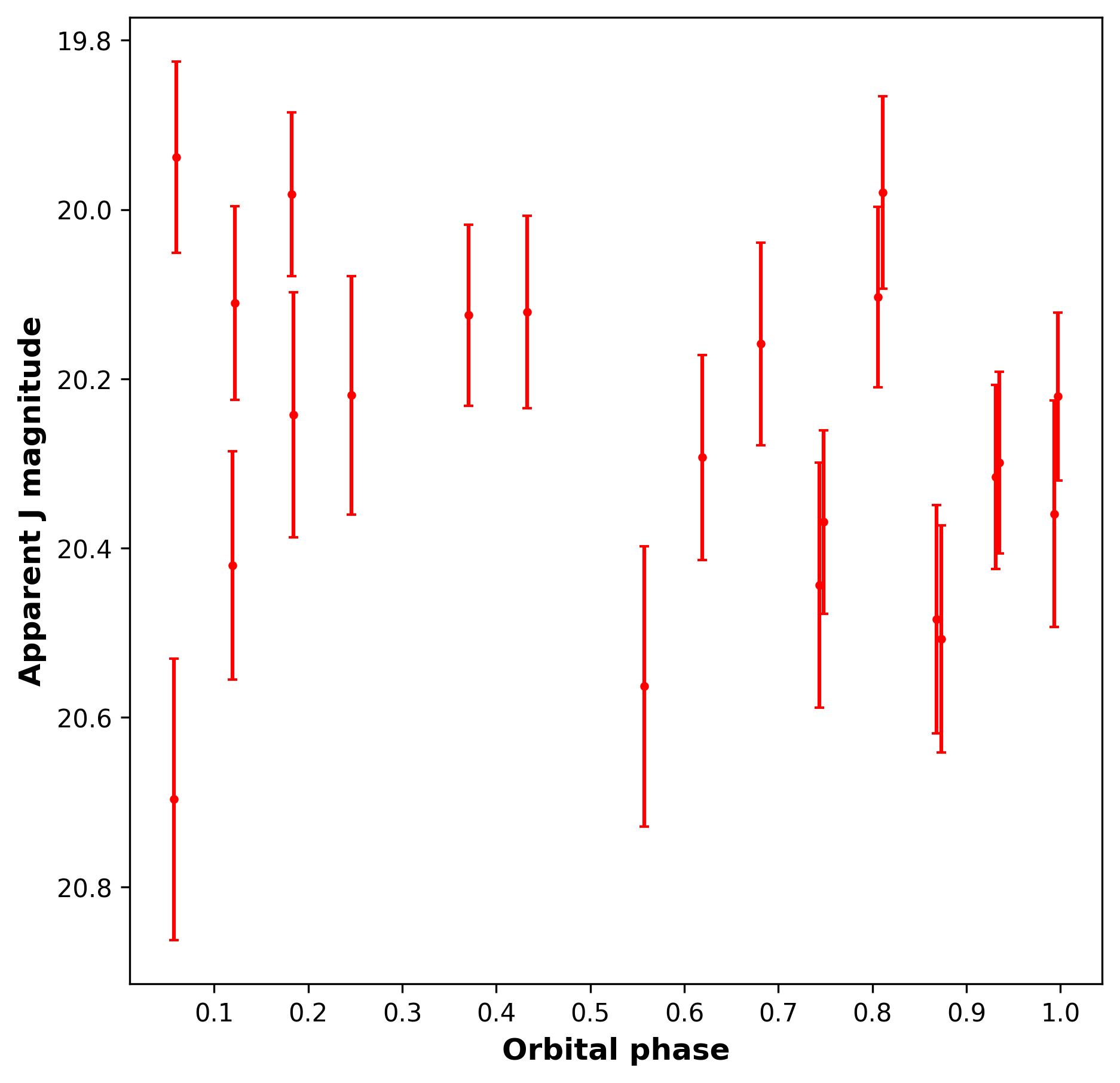}
    \caption{The $J$-band light curve shown in Fig. \ref{fig:IRlc} phase-folded to the orbital period from \citet{miao23}.}
    \label{fig:IRphase}
\end{figure}

Fig. \ref{fig:IRlc} shows the light curve of the nearest infrared source to the radio location of PSR J1720$-$0534 at the coordinates of RA: 260.2265 Dec.: -5.573. The measurements are consistent with a constant source with the average $J$-band magnitude of 20.25 and a scatter of 0.05 mag. The same light curve is folded with the known orbital period in Fig. \ref{fig:IRphase} with no evident orbital variability. 
 
\bsp	
\label{lastpage}
\end{document}